\documentclass[twocolumn]{aastex63}

\usepackage[T1]{fontenc}
\usepackage[utf8]{inputenc}
\usepackage[icelandic,english]{babel}
\usepackage{gensymb}
\usepackage{lmodern}
\usepackage{amsmath}
\usepackage{natbib}
\usepackage{bm}
\usepackage{fancyhdr}
\usepackage{amssymb}
\setcitestyle{citesep={;}}
\setcitestyle{aysep={}}

\newcommand\tess{TESS }
\newcommand{\RNum}[1]{\uppercase\expandafter{{\scshape\romannumeral #1\relax}}}

\shorttitle{TOI-1670 b and c}
\accepted{March 1, 2022}
\shortauthors{Tran et al.}
\graphicspath{{./}{Figures/}}

\begin{document}

\title{TOI-1670 b and c: An Inner Sub-Neptune with an Outer Warm Jupiter \\ Unlikely to have Originated from High-Eccentricity Migration}

\correspondingauthor{Quang H. Tran}
\author[0000-0001-6532-6755]{Quang H. Tran}
\email{quangtran@utexas.edu}
\affiliation{Department of Astronomy, The University of Texas at Austin, 2515 Speedway, Stop C1400, Austin, TX 78712, USA}

\author[0000-0003-2649-2288]{Brendan P. Bowler}
\affiliation{Department of Astronomy, The University of Texas at Austin, 2515 Speedway, Stop C1400, Austin, TX 78712, USA}

\author{Michael Endl}
\affiliation{Department of Astronomy, The University of Texas at Austin, 2515 Speedway, Stop C1400, Austin, TX 78712, USA}
\affiliation{McDonald Observatory, The University of Texas at Austin, 2515 Speedway, Stop C1400, Austin, TX 78712, USA}

\author[0000-0001-9662-3496]{William D. Cochran}
\affiliation{Department of Astronomy, The University of Texas at Austin, 2515 Speedway, Stop C1400, Austin, TX 78712, USA}
\affiliation{McDonald Observatory, The University of Texas at Austin, 2515 Speedway, Stop C1400, Austin, TX 78712, USA}

\author{Phillip J. MacQueen}
\affiliation{McDonald Observatory, The University of Texas at Austin, 2515 Speedway, Stop C1400, Austin, TX 78712, USA}

\author[0000-0001-8627-9628]{Davide Gandolfi}
\affiliation{Dipartimento di Fisica, Universit\`a degli Studi di Torino, via Pietro Giuria 1, I-10125, Torino, Italy}

\author[0000-0003-1257-5146]{Carina M. Persson}
\affiliation{Department of Space, Earth and Environment, Chalmers University of Technology, Onsala Space Observatory, 439 92 Onsala, Sweden}

\author[0000-0002-0855-8426]{Malcolm Fridlund}
\affiliation{Department of Space, Earth and Environment, Chalmers University of Technology, Onsala Space Observatory, 439 92 Onsala, Sweden}
\affiliation{Leiden Observatory, Leiden University, NL-2333 CA Leiden, The Netherlands}

\author[0000-0003-0987-1593]{Enric Palle}
\affiliation{Instituto de Astrof\'\i sica de Canarias (IAC), E-38205 La Laguna, Tenerife, Spain}
\affiliation{Departamento de Astrof\'\i sica, Universidad de La Laguna (ULL), E-38206, La Laguna, Tenerife, Spain}

\author[0000-0002-7031-7754]{Grzegorz Nowak}
\affiliation{Instituto de Astrof\'\i sica de Canarias (IAC), E-38205 La Laguna, Tenerife, Spain}
\affiliation{Departamento de Astrof\'\i sica, Universidad de La Laguna (ULL), E-38206, La Laguna, Tenerife, Spain}

\author[0000-0003-0047-4241]{Hans~J.~Deeg}
\affiliation{Instituto de Astrof\'\i sica de Canarias (IAC), E-38205 La Laguna, Tenerife, Spain}
\affiliation{Departamento de Astrof\'\i sica, Universidad de La Laguna (ULL), E-38206, La Laguna, Tenerife, Spain}

\author[0000-0002-4671-2957]{Rafael Luque}
\affiliation{Instituto de Astrof\'isica de Andaluc\'ia (IAA-CSIC), Glorieta de la Astronom\'ia s/n, 18008 Granada, Spain}

\author[0000-0002-4881-3620]{John H. Livingston}
\affiliation{Department of Astronomy, University of Tokyo, 7-3-1 Hongo, Bunkyo-ku, Tokyo 113-0033, Japan}
\affiliation{Astrobiology Center, 2-21-1 Osawa, Mitaka, Tokyo 181-8588, Japan}
\affiliation{National Astronomical Observatory of Japan, NINS, 2-21-1 Osawa, Mitaka, Tokyo 181-8588, Japan}

\author[0000-0002-1623-5352]{Petr Kab\'{a}th}
\affiliation{Astronomical Institute of the Czech Academy of Sciences, Fri\v{c}ova 298, 25165, Ond\v{r}ejov, Czech Republic}

\author[0000-0002-7602-0046]{Marek Skarka}
\affiliation{Astronomical Institute of the Czech Academy of Sciences, Fri\v{c}ova 298, 25165, Ond\v{r}ejov, Czech Republic}
\affiliation{Department of Theoretical Physics and Astrophysics, Masaryk University, Kotl\'{a}rsk\'{a} 2, CZ-61137, Brno, Czech Republic}

\author[0000-0002-5313-9722]{J\'an \v{S}ubjak}
\affiliation{Astronomical Institute of the Czech Academy of Sciences, Fri\v{c}ova 298, 25165, Ond\v{r}ejov, Czech Republic}
\affiliation{Astronomical Institute of Charles University, V Hole\v{s}ovi{\v c}k\'ach 2, 180 00, Praha, Czech Republic}

\author[0000-0002-2532-2853]{Steve B. Howell}
\affiliation{NASA Ames Research Center, Moffett Field, CA 94035, USA}

\author[0000-0003-1762-8235]{Simon H.\ Albrecht}
\affiliation{Stellar Astrophysics Centre, Department of Physics and Astronomy, Aarhus University, Ny Munkegade 120, DK-8000 Aarhus C, Denmark}

\author[0000-0001-6588-9574]{Karen A. Collins}
\affiliation{Center for Astrophysics | Harvard \& Smithsonian, 60 Garden
Street, Cambridge, MA 02138, USA}

\author{Massimiliano Esposito}
\affiliation{Th\"uringer Landessternwarte Tautenburg, Sternwarte 5, D-07778 Tautenburg, Germany}

\author[0000-0001-5542-8870]{Vincent Van Eylen}
\affiliation{Mullard Space Science Laboratory, University College London, Holmbury St Mary, Dorking, Surrey RH5 6NT, UK}

\author[0000-0003-3370-4058]{Sascha Grziwa}
\affiliation{Rheinisches Institut für Umweltforschung an der Universiät zu Köln, Aachener Straße 209, D-50931 Köln, Germany}

\author[0000-0001-9670-961X]{Elisa Goffo}
\affiliation{Dipartimento di Fisica, Universit\`a degli Studi di Torino, via Pietro Giuria 1, I-10125, Torino, Italy}
\affiliation{Th\"uringer Landessternwarte Tautenburg, Sternwarte 5, D-07778 Tautenburg, Germany}

\author[0000-0003-0918-7484]{Chelsea~ X.~Huang}
\affiliation{Department of Physics and Kavli Institute for Astrophysics and Space Research, Massachusetts Institute of Technology, Cambridge, MA 02139, USA}
\affiliation{Juan Carlos Torres Fellow}

\author[0000-0002-4715-9460]{Jon~M.~Jenkins}
\affiliation{NASA Ames Research Center, Moffett Field, CA 94035, USA}

\author[0000-0003-0751-3231]{Marie Karjalainen}
\affiliation{Astronomical Institute of the Czech Academy of Sciences, Fri\v{c}ova 298, 25165, Ond\v{r}ejov, Czech Republic}

\author[0000-0002-2656-909X]{Raine Karjalainen}
\affiliation{Astronomical Institute of the Czech Academy of Sciences, Fri\v{c}ova 298, 25165, Ond\v{r}ejov, Czech Republic}

\author[0000-0001-7880-594X]{Emil Knudstrup}
\affiliation{Stellar Astrophysics Centre, Department of Physics and Astronomy, Aarhus University, Ny Munkegade 120, DK-8000 Aarhus C, Denmark}

\author[0000-0002-0076-6239]{Judith Korth}
\affiliation{Department of Space, Earth and Environment, Astronomy and Plasma Physics, Chalmers University of Technology, 412 96 Gothenburg, Sweden}

\author[0000-0002-9910-6088]{Kristine W. F. Lam}
\affiliation{Institute of Planetary Research, German Aerospace Center (DLR), Rutherfordstra{\ss}e 2, 12489 Berlin, Germany}

\author[0000-0001-9911-7388]{David W. Latham}
\affiliation{Center for Astrophysics | Harvard \& Smithsonian, 60 Garden
Street, Cambridge, MA 02138, USA}

\author[0000-0001-8172-0453]{Alan M. Levine}
\affiliation{Department of Physics and Kavli Institute for Astrophysics and Space Research, Massachusetts Institute of Technology, Cambridge, MA 02139, USA}

\author[0000-0002-4143-4767]{H. L. M. Osborne}
\affiliation{Mullard Space Science Laboratory, University College London, Holmbury St Mary, Dorking, Surrey RH5 6NT, UK}

\author[0000-0002-8964-8377]{Samuel N. Quinn}
\affiliation{Center for Astrophysics | Harvard \& Smithsonian, 60 Garden Street, Cambridge, MA 02138, USA}

\author[0000-0003-3786-3486]{Seth Redfield}
\affiliation{Astronomy Department and Van Vleck Observatory, Wesleyan University, Middletown, CT 06459, USA}

\author[0000-0003-2058-6662]{George~R.~Ricker}
\affiliation{Department of Physics and Kavli Institute for Astrophysics and Space Research, Massachusetts Institute of Technology, Cambridge, MA 02139, USA}

\author[0000-0002-6892-6948]{S.~Seager}
\affiliation{Department of Physics and Kavli Institute for Astrophysics and Space Research, Massachusetts Institute of Technology, Cambridge, MA 02139, USA}
\affiliation{Department of Earth, Atmospheric and Planetary Sciences, Massachusetts Institute of Technology, Cambridge, MA 02139, USA}
\affiliation{Department of Aeronautics and Astronautics, Massachusetts Institute of Technology, 77 Massachusetts Avenue, Cambridge, MA 02139, USA}

\author{Luisa Maria Serrano}
\affiliation{Dipartimento di Fisica, Universit\`a degli Studi di Torino, via Pietro Giuria 1, I-10125, Torino, Italy}

\author[0000-0002-2386-4341]{Alexis M. S. Smith}
\affiliation{Institute of Planetary Research, German Aerospace Center (DLR), Rutherfordstra{\ss}e 2, 12489 Berlin, Germany}

\author[0000-0002-6778-7552]{Joseph D. Twicken}
\affiliation{NASA Ames Research Center, Moffett Field, CA 94035, USA}
\affiliation{SETI Institute, Mountain View, CA  94043, USA}

\author[0000-0002-4265-047X]{Joshua N.\ Winn}
\affiliation{Department of Astrophysical Sciences, Peyton Hall, 4 Ivy Lane, Princeton, NJ 08544, USA}

\begin{abstract}
    We report the discovery of two transiting planets around the bright ($V=9.9$ mag) main sequence F7 star TOI-1670 by the \textit{Transiting Exoplanet Survey Satellite}. TOI-1670 b is a sub-Neptune ($R_\mathrm{b} = 2.06_{-0.15}^{+0.19}$ $R_\earth$) on a 10.9-day orbit and TOI-1670 c is a warm Jupiter ($R_\mathrm{c} = 0.987_{-0.025}^{+0.025}$ $R_\mathrm{Jup}$) on a 40.7-day orbit. Using radial velocity observations gathered with the Tull coud\'e Spectrograph on the Harlan J. Smith telescope and HARPS-N on the Telescopio Nazionale Galileo, we find a planet mass of $M_\mathrm{c} = 0.63_{-0.08}^{+0.09}$ $M_\mathrm{Jup}$ for the outer warm Jupiter, implying a mean density of $\rho_c = 0.81_{-0.11}^{+0.13}$ g cm$^{-3}$. The inner sub-Neptune is undetected in our radial velocity data ($M_\mathrm{b} < 0.13$ $M_\mathrm{Jup}$ at the 99\% confidence level). Multi-planet systems like TOI-1670 hosting an outer warm Jupiter on a nearly circular orbit ($e_\mathrm{c} = 0.09_{-0.04}^{+0.05}$) and one or more inner coplanar planets are more consistent with ``gentle'' formation mechanisms such as disk migration or \textit{in situ} formation rather than high-eccentricity migration. Of the 11 known systems with a warm Jupiter and a smaller inner companion, 8 (73\%) are near a low-order mean-motion resonance, which can be a signature of migration. TOI-1670 joins two other systems (27\% of this subsample) with period commensurabilities greater than 3, a common feature of \textit{in situ} formation or halted inward migration. TOI-1670 and the handful of similar systems support a diversity of formation pathways for warm Jupiters.
\end{abstract}

\keywords{planetary systems, planets and satellites: detection, stars: individual (TOI-1670)}

\section{\textbf{Introduction}}
\label{sec:intro}
    
    The origin of giant planets interior to the water ice line remains an open question. A number of theories have been proposed to explain the closest-in ($P < 10$ d) giant planets, or hot Jupiters \citep[HJs; e.g.,][]{Dawson2018, Fortney2021}. These scenarios are primarily divided between dynamically ``violent'' or ``gentle'' mechanisms. The former consists of three-body dynamical interactions such as planet-planet scattering or high-eccentricity tidal migration \citep[e.g.,][]{Wu2003, Fabrycky2007, Triaud2010, Naoz2011, Batygin2012}. The latter refers to disk migration \citep[e.g.,][]{Ward1997, Albrecht2012, Kley2012} or \textit{in situ} formation \citep[e.g.,][]{Boley2016,Huang2016, Batygin2016, Anderson2020}. These processes have also been used to explain part of the farther-out population of warm Jupiters (WJs; defined here to have $10 < P < 200$ d). However, observed WJ demographics suggest that multiple processes are present in sculpting these more distant giant systems.
    
    WJs can be broadly divided into two classes. The first is a transient population that will likely evolve into HJs. In more disruptive formation mechanisms such as high-eccentricity tidal migration, giant planets at comparatively wide separations are disturbed onto highly eccentric orbits by a third body via planet-planet scattering or Von Zeipel–Lidov–Kozai oscillations and eventually circularize into orbits with shorter periods \citep{Lidov1962, Kozai1962, Naoz2016, Ito2019}. Eccentric giant planets undergoing this tidally damped inward migration are caught in a rapid, temporary state \citep{Naef2001, Dawson2018, Dong2021, Jackson2021}. They are expected to start their journeys with much higher eccentricities \citep[$e \gtrsim 0.9$;][]{Vick2019} which can decay as rapidly as $\sim$1 Myr as they settle in near their host star \citep{Patra2020, Mancini2021}. The majority of WJs are not expected to belong to this transient classification.
    
    Instead, most WJs are a part of a ``static'' population that will remain stable over long time periods. This group consists of the apparently single systems with low to moderate eccentricities as well as co-planar multi-planet systems containing WJs with low eccentricities. These WJs have periapses larger than what is required for efficient tidal damping of their orbits, which occurs at $\lesssim$0.05 AU \citep{Anderson2016, Dong2021}, so these planets cannot be undergoing high-eccentricity migration. If most giant planets form beyond the water ice line, other migration mechanisms must play a major role in sculpting these WJ orbital properties and demographics \citep[e.g.,][]{Veras2005, Fogg2009, Dong2014, Ortiz2015, Huang2016, Anderson2017, Anderson2020, Schlecker2020}. However, the relative importance of these pathways is still unknown. Investigating WJ orbital eccentricities can place additional constraints on the dominant giant planet migration mechanism since each scenario will produce different observed eccentricity distributions.
    
    Warm Jupiters have an eccentricity distribution that peaks at $e = 0.0$ with a tail that extends out to $e \sim 0.8$ \citep{Kipping2013, Dong2021}. In order to produce the population of WJs with moderately eccentric orbits ($e \sim 0.2-0.7$), a mechanism is needed that can excite eccentricities. These potential excitation scenarios include interactions involving a disk \citep[e.g.,][]{Goldreich2003, Petrovich2019}, secular eccentricity oscillations driven by interactions with a distant inclined giant planet \citep[e.g.,][]{Anderson2017}, and planet-planet scattering events \citep[e.g.,][]{Mustill2017, Frelikh2019, Marzari2019, Anderson2020}. An important clue is the observed dependence on metallicity of the giant planet eccentricity distribution, where metal-rich systems (that may more favorably form multiple giant planets) are more likely to host eccentric gas giants \citep{Dawson2013}.
    
    Systems hosting WJs with low-mass inner companions on coplanar orbits are especially useful laboratories to test these planet formation and migration theories. Their small orbital eccentricities and low mutual inclinations suggest that disk migration or \textit{in situ} formation likely helped create this population. WJs have a relatively high close companion rate of nearly 50\% \citep{Huang2016}. However, their intrinsically low occurrence rate \citep[$\sim$1--2\%;][]{Cumming2008} combined with the difficulty of detecting lower-mass inner planets means that only a handful of known multi-planet systems host a WJ \citep{Johnson2010, Santerne2016, Fernandes2019}. Increasing the number of systems with this multi-planet architecture may further distinguish this subsample into two WJ populations, each of which likely reflects different formation and migration routes.
    
    Here we present the discovery of the transiting multi-planet system TOI-1670 bc, a warm Jupiter (TOI-1670 c) with an inner sub-Neptune (TOI-1670 b) found with the \textit{Transiting Exoplanet Survey Satellite} \citep[TESS;][]{Ricker2015}. TOI-1670 b and c were originally identified by the \tess Science Processing Operations Center \citep[SPOC;][]{Jenkins2016} pipeline as two promising transiting signals that were subsequently promoted to TESS Object of Interest \citep[TOI;][]{Guerrero2021} status. TOI-1670 (TIC ID 441739020; 2MASS J17160415+7209402; \textit{Gaia} DR2 1651911084230149248) is a relatively inactive (log$R^\prime_\mathrm{HK} = -4.93$) old F7 dwarf with a \tess apparent magnitude of 9.5 mag and a moderate projected rotational velocity of $\approx$9 km s$^{-1}$ (see \autoref{sec:analysis}). In this work, we validate both planets and measure the mass of the outer planet TOI-1670 c. In \autoref{sec:observations}, we describe the \tess photometric data and follow-up radial velocity (RV) observations used in the planet validation and mass measurement. Our characterization of the system, including the host star and a global fit to the RVs and light curve, is presented in \autoref{sec:analysis}. We conclude in \autoref{sec:discuss} by contextualizing TOI-1670 in the paradigm of WJs and their formation.

\section{\textbf{Observations}}
\label{sec:observations}

    KESPRINT\footnote{ \href{http://kesprint.science/}{http://kesprint.science/}} is an international collaboration focused on the discovery, confirmation, and characterization of exoplanet candidates from space-based missions \citep[e.g.,][]{Persson2018, Gandolfi2018, Livingston2019, Lam2020, Subjak2020}. As part of this consortium, a series of ground-based follow-up observations of TOI-1670 were taken. These data are primarily used to reject the possibility of a false positive scenario in which the observed transiting signal is caused by something other than a planet. For example, this includes a low-mass eclipsing binary (EB), a grazing transit of an EB, a background EB, or a transiting planet around a background star. Reconnaissance spectra are used to exclude an EB scenario by constraining the maximum amplitude of the RV signal. High-resolution speckle images are taken to exclude binary companions to TOI-1670 and nearby background stars. High-resolution spectra are used to characterize the host star and, when possible, measure the masses of the planets.

\subsection{TESS Photometry}
\label{sec:TESS_phot}
    TOI-1670 was observed by \tess at 2-minute cadence over 11 sectors (15, 16, 18, 19, 20, 21, 22, 23, 24, 25, and 26) for a total of 323 days. Images were reduced and light curves were analyzed for transit signals with the \tess SPOC pipeline \citep{Jenkins2016}, which identified two potential transit signals \citep{Jenkins2002, Jenkins2010, Jenkins2020} with periods of 40.7 days (TOI 1670.01) and 10.9 days (TOI 1670.02). SPOC vetting tests \citep{Twicken2018, Li2019} validated both signals as consistent with planets and they were designated as TOIs \citep{Guerrero2021} by the TESS Science Office.
    
    \begin{figure*}[!ht]
	    \centering
	    \includegraphics[width=1.0\linewidth]{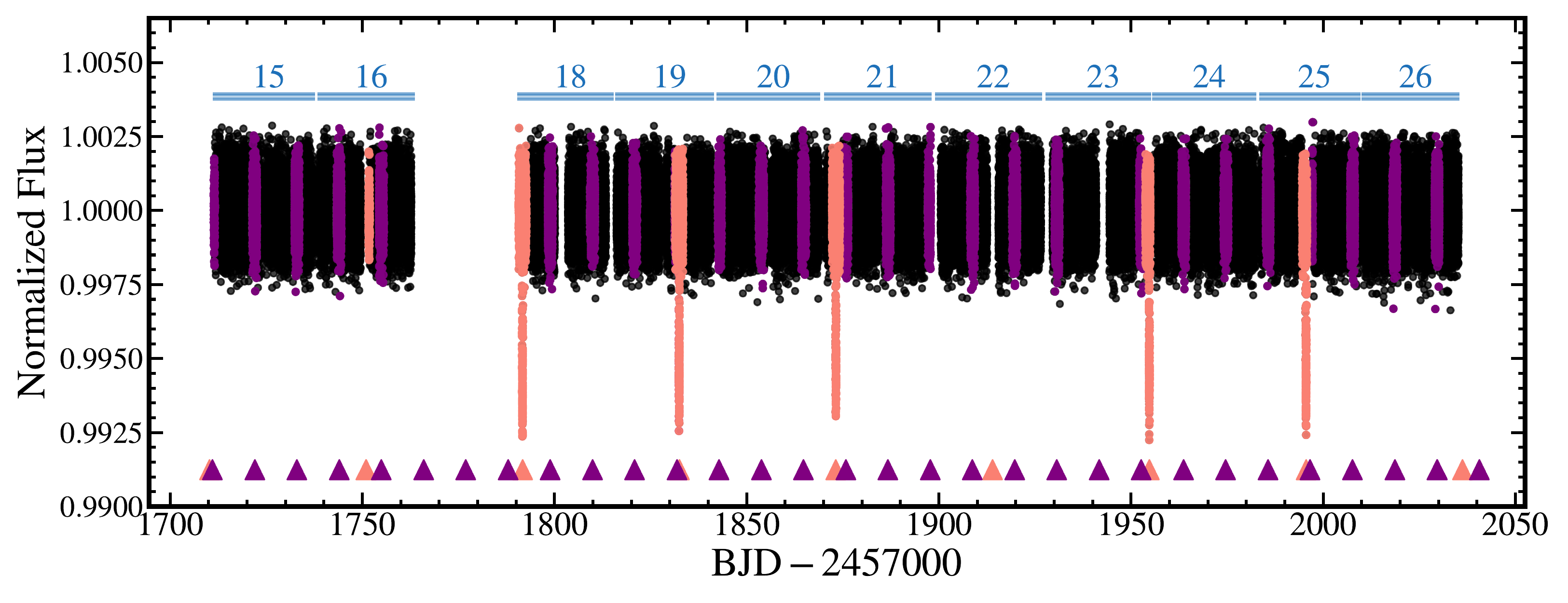}
	    \caption{The detrended \tess light curve of TOI-1670. The full light curve is shown in black. Colored purple and pink points are relative photometry within four transit duration before and after transits of TOI-1670 b and c, respectively, and are used in the global RV and light curve model fit. Times of transit are also marked by triangles plotted along the time axis. Gaps in the light curve correspond to periods where \tess uploaded data, result from data quality cuts, or are during Sector 17 when TOI-1670 was not observed. \tess sectors are shown in blue above the light curve.}
	    \label{fig:lightcurve}
    \end{figure*}
    
    We downloaded the SPOC Pre-search Data Conditioning Simple Aperture Photometry (PDCSAP) light curve \citep{Smith2012, Stumpe2012, Stumpe2014} from the MAST data archive\footnote{\href{https://archive.stsci.edu/missions-and-data/tess}{https://archive.stsci.edu/missions-and-data/tess/}} using the \texttt{lightkurve} \citep{Lightkurve2018} software package. We removed all of the photometric measurements that are flagged as poor quality by the SPOC pipeline (DQUALITY > 0) or where either the flux or flux error is listed as \texttt{NaN}. Outlier rejection was performed at 3$\sigma$ for positive outliers and 10$\sigma$ for negative outliers to allow for transit events. The lightcurve is flattened by removing low frequency trends using a Savitzky-Golay filter \citep{Savitzky1964} after all transit events were masked out. The final light curve for TOI-1670 is shown in \autoref{fig:lightcurve}. Photometric points used in the global model fit are colored purple and pink. These cover the transit events for TOI-1670 b and c, respectively, and their times of transit are further denoted by the corresponding colored triangles along the time axis.

\subsection{TRES Reconnaissance Spectroscopy}
    We obtained six spectra of TOI-1670 with the Tillinghast Reflector \'Echelle Spectrograph \citep[TRES;][]{Furesz2008} on the 1.5-meter Tillinghast telescope at the Fred L. Whipple Observatory on UT 2020 February 2 and 20, UT 2020 March 6, 9, and 16, and UT 2020 July 7. Exposure times range from 300 to 650 seconds and have an average S/N of $32\pm5$. Radial velocities were extracted following \citet{Buchhave2010}. Spectra have an average measurement error of 53 m s$^{-1}$ and an RMS of 54 m s$^{-1}$, which exclude the possibility of an eclipsing binary scenario; however these spectra are not used as part of the orbit fit. \autoref{tab:recon_RVs} in \autoref{sec:appendix_rvs} reports the RV measurements.
    
\subsection{OES Reconnaissance Spectroscopy}
\label{sec:Ondrejov_spec}
    We collected 32 spectra using the Ond\v rejov \'Echelle Spectrograph (OES) on the 2-meter Perek telescope at the Ond\v rejov Observatory in the Czech Republic \citep{Kabath2020}. These observations were obtained between UT 2020 February and UT 2020 September at a cadence of $3-5$ RVs per month. We extracted the spectra and performed the bias, flat-field, and cosmic ray corrections using standard IRAF 2.16 routines \citep{Tody1993}. RVs were extracted using the IRAF cross correlation \texttt{fxcor} taking the highest S/N spectrum as a template. The average measurement error is 110 m s$^{-1}$ and the RV RMS is 116 m s$^{-1}$. The Doppler signals for TOI-1670 b and c are not detected in this dataset so they are not used in the orbit fit. However, they are used to reject an eclipsing binary scenario and justified further follow-up of TOI-1670 with precise RV measurements. The reconnaissance RV measurements are reported in \autoref{tab:recon_RVs} in \autoref{sec:appendix_rvs}.

\subsection{Tull coud\'e Spectroscopy}
\label{sec:Tull_spec}
    We used the Tull coud\'e Spectrograph on the 2.7-m Harlan J. Smith telescope at McDonald Observatory to obtain 49 spectra of TOI-1670 between UT 2020 April and UT 2021 September. The Tull coud\'e Spectrograph is a cross-dispersed \'echelle spectrograph with a wavelength coverage ranging from 3750 $\AA$ to 10200 $\AA$ \citep{Tull1995}. Our configuration uses a $1.2\arcsec{}$ slit which yields a resolving power of $R=$ 60,000. Precise wavelength calibration and instrumental profile reconstruction is achieved with a temperature-controlled iodine vapor (I$_2$) cell that is mounted in front of the entrance slit.

    Radial velocities are extracted using the RV reduction pipeline \texttt{Austral} \citep{Endl2000}. The I$_2$ cell imprints a well-understood reference absorption spectrum onto the stellar spectra. Precise differential radial velocities are then calculated by comparing each stellar-plus-iodine spectrum with a high S/N stellar template devoid of iodine lines. The \textit{S}-index activity metric for each spectrum is also calculated and calibrated onto the Mt. Wilson \textit{S}-index system following the description in \citet{Paulson2002}. \autoref{tab:RVs} in \autoref{sec:appendix_rvs} reports the resulting RVs, activity indices, and related measurement errors.

\subsection{FIES Spectroscopy}
\label{fies_spectra}
    We acquired 7 spectra of TOI-1670 using the Fibre-fed {\'E}chelle Spectrograph \citep[FIES;][]{1999anot.conf...71F,2014AN....335...41T} at the 2.56-m Nordic Optical Telescope \citep[NOT;][]{Djupvik2010} of Roque de los Muchachos Observatory (La Palma, Spain). The observations were carried out between UT 2020 May 25 and UT 2020 September 6 as part of the Spanish CAT observing program 59-210. We used the FIES high-resolution mode, which provides a resolving power of $R=$ 67,000 in the spectral range $3760-8220$ $\AA$. We traced the RV drift of the instrument by acquiring long-exposed ThAr spectra (exposure time of 90 seconds) immediately before and after each science observation. The science exposure time was set to $1200-1800$ seconds, depending on the sky conditions and scheduling constraints. The data reduction follows the steps described in \cite{2010ApJ...720.1118B} and \cite{2015A&A...576A..11G} and includes bias subtraction, flat fielding, order tracing and extraction, and wavelength calibration. Radial velocities were derived via multi-order cross-correlations, using the first stellar spectrum as a template. The SNR per pixel at 5500~\AA\ ranges between 40 and 65. The average RV uncertainty is $13.4 \pm 2.4$ m s$^{-1}$.
    
\subsection{HARPS-N Spectroscopy}
\label{sec:HARPS-N_spec}
    We observed TOI-1670 with the HARPS-N spectrograph ($R\approx$ 115,000) on the 3.59-m Telescopio Nazionale Galileo (TNG) at Roque de los Muchachos Observatory located in La Palma, Spain between UT 2020 August and UT 2020 September \citep{Cosentino2012, Cosentino2014} during observing program A40TAC\_22 (PI: Gandolfi). A total of 8 spectra were taken; seven spectra had an exposure time of 1800 seconds and one had an exposure time of 215 seconds. This resulted in an average S/N at 550 nm of $84\pm16$ for the first seven spectra and a S/N of 15 for the shorter exposure.
    
    We used the standard HARPS-N Data Reduction Software (DRS) with a G2 numerical mask to extract the RVs \citep{Pepe2002}. The RVs, their measurement errors, and associated activity indicators such as the bisector inverse slope (BIS), FWHM of the cross-correlation function (CCF), and \textit{S}-index produced by the HARPS-N DRS are listed in \autoref{tab:RVs} of \autoref{sec:appendix_rvs}.
    
\subsection{High-resolution Imaging}

\begin{figure}[!t]
	    \centerline{\includegraphics[width=1.08\linewidth]{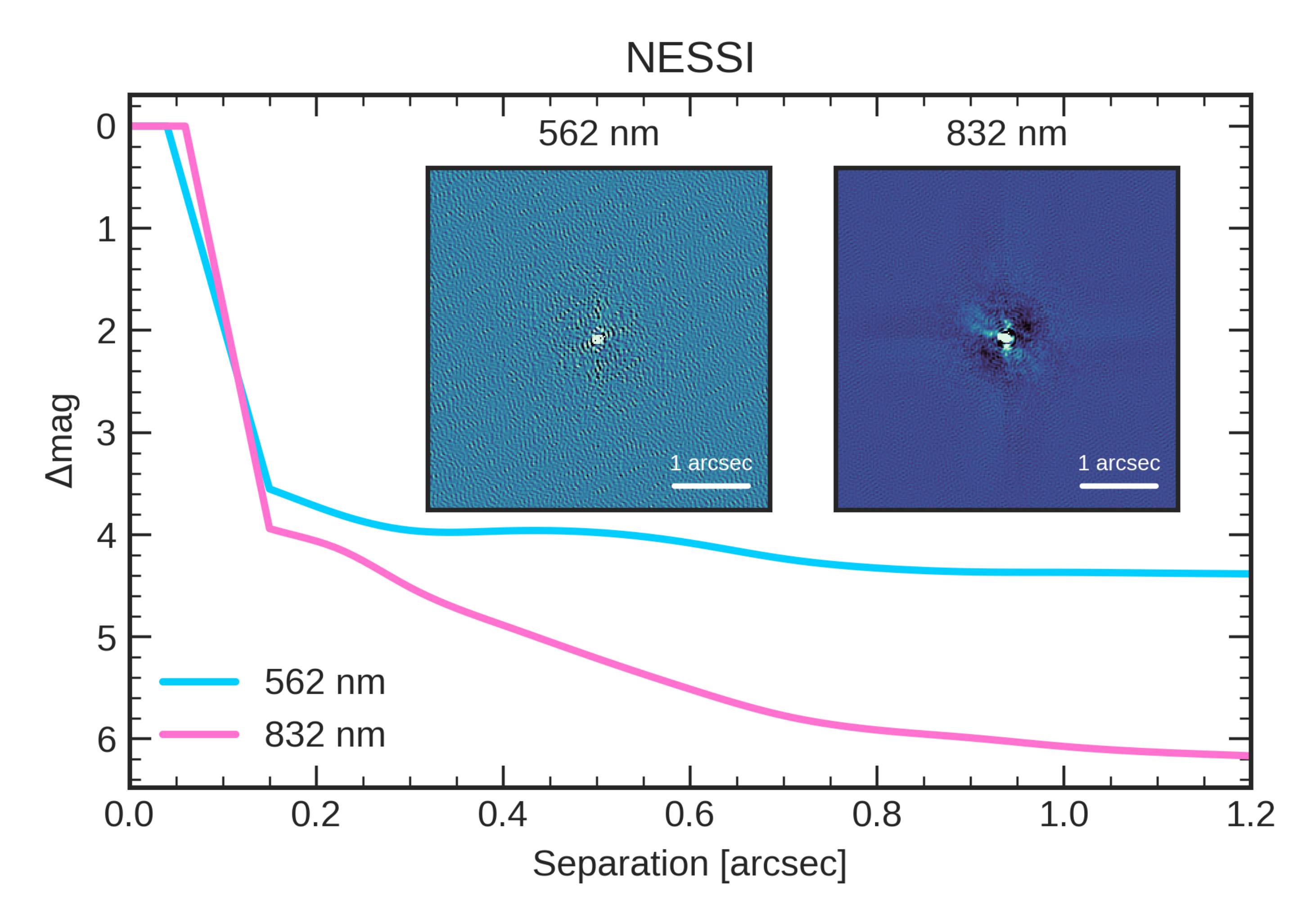}}
  	    \centerline{\includegraphics[width=1.08\linewidth]{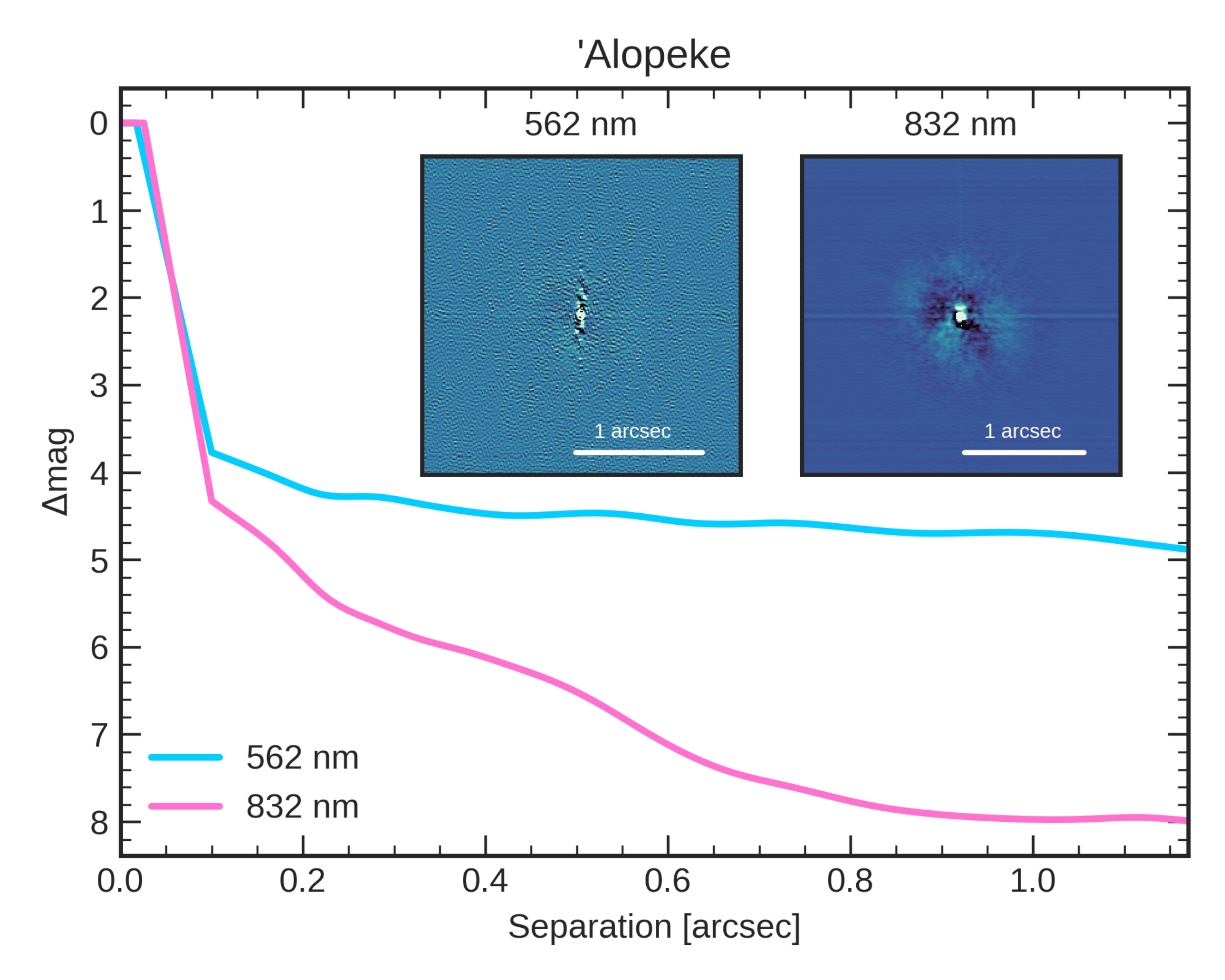}}
	    \caption{Reconstructed speckle images of TOI-1670 from NESSI (top) and `Alopeke (bottom) in the 562 nm and 832 nm bands and their corresponding $5\sigma$ contrast curves. North is up and east is to the left.}
    \label{fig:speckle_images}
    \end{figure}

    On the nights of UT 2021 April 5 and June 24, TOI-1670 was observed with the NESSI and `Alopeke speckle imagers \citep{2018ScottNESSI, Scott2019}, mounted on the 3.5-m WIYN telescope at Kitt Peak and the 8.1-m Gemini North telescope on Mauna Kea, respectively. Both instruments simultaneously acquire data in two bands centered at 562 nm and 832 nm using high speed electron-multiplying CCDs (EMCCDs). Observations of TOI-1670 were performed in the 562 nm and 832 nm bands following the procedures described in \citet{Howell2011}. The resulting reconstructed images have a 5$\sigma$ delta magnitude contrast of 4 to 8 magnitudes at angular separations from 20 mas to 1.2\arcsec{} in the 832 nm band (\autoref{fig:speckle_images}). No other companion sources are detected in the reconstructed images within these angular limits down to the contrasts obtained. These angular limits correspond to spatial separations of 3.3 to 200 AU at the distance of TOI-1670.
    
\clearpage
\section{\textbf{Analysis}}
\label{sec:analysis}

\subsection{Stellar Parameters}
\label{sec:star_params}

    Planetary parameters measured from the global joint fit of the transit and RV data depend on precise stellar mass and radius measurements. In particular, $R_p$ is determined from the transit depth and $M_p$ from the RV semi-amplitude, which are dependent on $R_*$ and $M_*$, respectively. The stellar mass and radius can be inferred with atmospheric and evolutionary models using the spectroscopic parameters ($T_\mathrm{eff}$, log$g$, [Fe/H], $v$sin$i$). We determine both these spectroscopic and fundamental parameters for TOI-1670 using several approaches described below.
    
    \subsubsection{Spectral Analysis}\label{sec:spec_analysis}
    
    We analyzed the co-added HARPS-N (S/N = 180) spectrum with the spectral analysis package Spectroscopy Made Easy \citep[\texttt{SME;}][]{Valenti1996, Valenti2005, Piskunov2017}. \texttt{SME}'s spectral fitting technique minimizes the $\chi^2$ value by fitting synthetic spectra of stars based on grids of atmosphere models and observations. We fit the co-added HARPS-N spectrum with the \texttt{ATLAS12} model spectra \citep{Kurucz2013} using the non-local thermodynamic equilibrium (non-LTE) \texttt{SME} version 5.2.2 following the procedure described in \citet{Fridlund2017} to compute $T_\mathrm{eff}$, log$g$, $v$sin$i$, and chemical abundances. The stellar surface gravity, log$g$, was estimated using the spectral wings of the Ca \RNum{1} 6102, 6122, 6162 \AA{} triplet and the Ca \RNum{1} 6439 \AA{} line. The microscopic and macroscopic turbulences, $V_\mathrm{mic}$ and $V_\mathrm{mac}$, were held fixed to the values determined in the calibration for stars with similar $T_\mathrm{eff}$ and log$g$ from \citet{Bruntt2010} and \citet{Doyle2014}, respectively.
    
    We also derive the stellar parameters using the publicly available \texttt{SpecMatch-Emp} software package \citep{Yee2017}. \texttt{SpecMatch-Emp} compares the HARPS-N template spectrum to a high resolution ($R \sim 55000$), high S/N ($>$100) Keck/HIRES optical spectral library of 404 well-characterized early- to late-type dwarfs (F1 to M5). The empirical spectra are calibrated using interferometry so \texttt{SpecMatch-Emp} produces estimates for $T_\mathrm{eff}$, [Fe/H], and $R_*$ (instead of log$g$). Prior to running the code, we convert the HARPS-N spectrum template onto the Keck/HIRES format following the procedure described in \citet{Hirano2018}.
    
    The stellar parameters derived from \texttt{SME} and \texttt{SpecMatch-Emp} are in good agreement with each other (\autoref{tab:spec_params}). From \texttt{SME}, we find an effective temperature of $T_\mathrm{eff} = 6170 \pm 61$ K and metallicity of [Fe/H] = 0.09 $\pm$ 0.07 dex, while \texttt{SpecMatch-Emp} gives $T_\mathrm{eff} = 6048 \pm 110$ K and [Fe/H] $= 0.05 \pm 0.09$ dex; these are consistent with each other within 1$\sigma$. These results are also in good agreement with the photometrically derived effective temperature from \textit{Gaia} DR2 ($T_\mathrm{eff} = 6162_{-175}^{+162}$ K) and agree at the 2$\sigma$ level with the TESS Input Catalog (TIC) v8 \citep{Guerrero2021} value of $6345 \pm 121$ K. For this work, we adopt the spectroscopic parameters from \texttt{SME} as it produces all atmospheric parameters. The final adopted stellar parameters are reported in \autoref{tab:star_params}.
    
    \begin{deluxetable*}{ccccc}[!ht]
    \setlength{\tabcolsep}{14pt}
    \tablecaption{Spectroscopic parameters of TOI-1670 derived using \texttt{SME} and \texttt{SpecMatch-Emp}. \label{tab:spec_params}}
    \tablehead{\colhead{Method} & \colhead{$T_\mathrm{eff}$ (K)} & \colhead{log$g$ (g cm$^{-3}$)} & \colhead{[Fe/H] (dex)} & \colhead{$v$sin$i$ (km s$^{-1}$)}}
    \startdata
    \texttt{SME} & $6170 \pm 61$ & $4.29 \pm 0.11$ & $0.09 \pm 0.07$ & $9.2 \pm 0.6$ \\
    \texttt{SpecMatch-Emp} & $6048 \pm 110$ & $\cdots$ & $0.05 \pm 0.09$ & $\cdots$ \\
    \enddata
    \end{deluxetable*}
    
    \subsubsection{Stellar Mass and Radius}
    \label{sec:mass_radius}
    
\begin{figure}[!t]
	    \centerline{\includegraphics[width=1.0\linewidth]{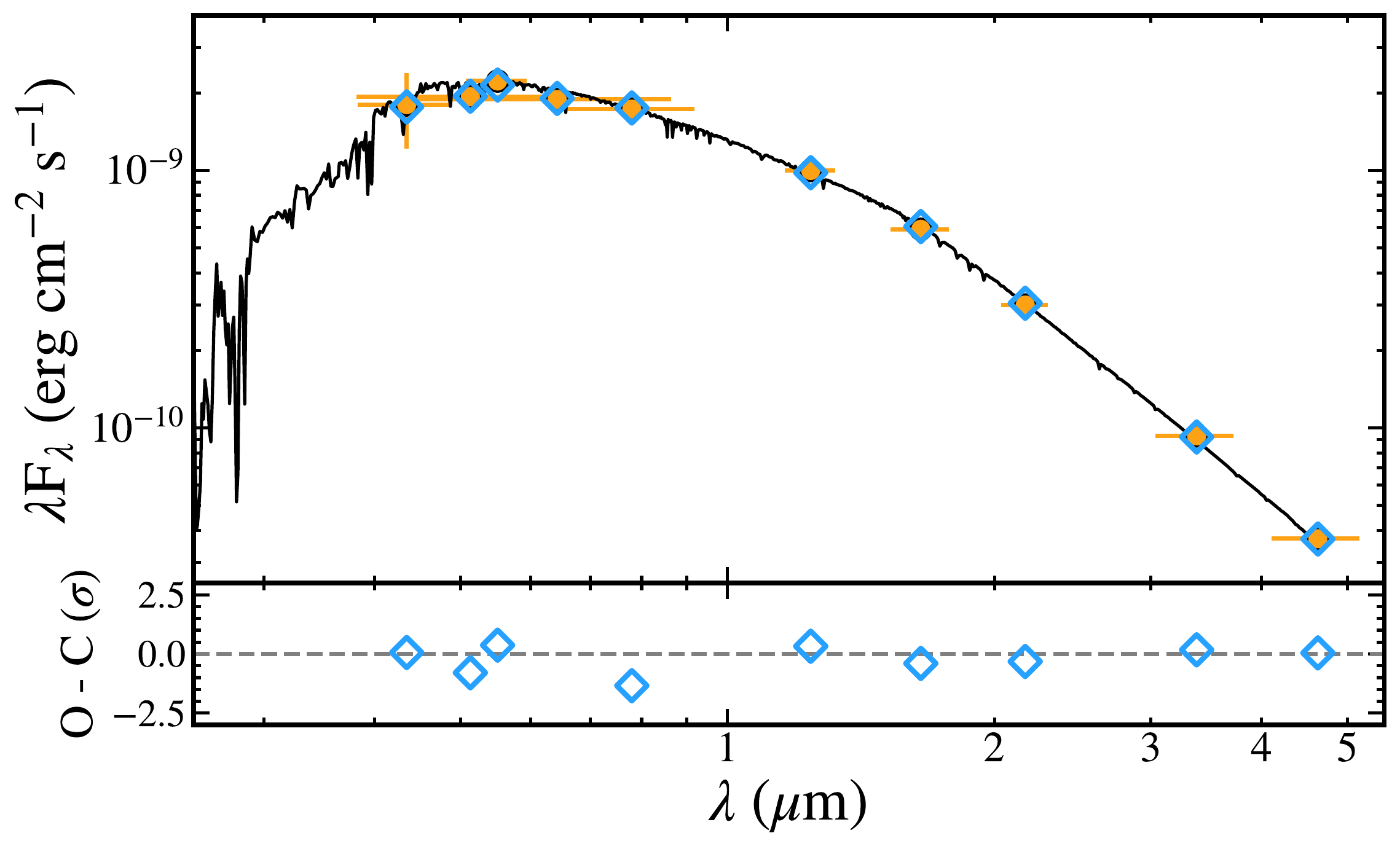}}
	    \caption{SED of TOI-1670. Broadband photometry (\autoref{tab:star_params}) is shown in orange circles, with horizontal errors representing the bandpass width. The best-fitting model SED is shown in black and blue diamonds are the model flux integrated over each bandpass. The residuals normalized by the photometric errors are shown in the bottom panel.}
    \label{fig:sed}
    \end{figure}

    We infer the stellar radius by fitting the spectral energy distribution (SED) of TOI-1670 using the software package \texttt{ARIADNE}\footnote{\href{https://github.com/jvines/astroARIADNE}{https://github.com/jvines/astroARIADNE}}. \texttt{ARIADNE} utilizes a Bayesian Model Averaging (BMA) framework that convolves four stellar atmosphere models---\texttt{Phoenix v2} \citep{Husser2013}, BT-Settl \citep{Allard2011}, \citet{Kurucz1993}, and \citet{Castelli2003}---with the response functions of commonly available broadband filters. For our SED fitting, we use the 2MASS $JHK_s$, \textit{Gaia} DR2 ($G$, $G_\mathrm{BP}$, $B_\mathrm{RP}$), Johnson $V$ and $B$, and WISE ($W1$ and $W2$) bandpasses. Synthetic SEDs are created by interpolating in $T_\mathrm{eff}$--$\mathrm{log}g$--$\mathrm{[Fe/H]}$ space. Distance, radius, $A_V$, and excess photometric uncertainty terms are free parameters in the fitting process. We set the priors for $T_\mathrm{eff}$, log$g$, and [Fe/H] to the values we found in \autoref{sec:spec_analysis}; the distance prior to the \citet{Bailer-Jones2021} Bayesian-based value ($165.72^{+0.32}_{-0.38}$ pc); and the stellar radius prior to the \textit{Gaia} DR2 value ($R_* = 1.38^{+0.08}_{-0.07}$ $R_\odot$). The extinction, $A_V$, has a flat prior limited by the maximum line-of-sight reddening according to the re-calibrated SFD galaxy dust map \citep{Schlegel1998, Schlafly2011}. The excess photometric noise parameters all have Gaussian priors centered at 0 with a standard deviation equal to 10 times the reported photometric error. The SED of TOI-1670 and best-fitting model are shown in \autoref{fig:sed}.
    
    We estimate the mass of TOI-1670 using the stellar isochrone software package \texttt{isochrones} \citep{Morton2015b} and the MESA Isochrones and Stellar Tracks \citep[MIST;][]{Dotter2016, Choi2016} evolutionary model grids. \texttt{isochrones} infers fundamental stellar parameters by comparing a variety of observational inputs to interpolated model values. We input the \textit{Gaia} DR2 parallax, broadband photometry (2MASS $JHK_s$; \textit{Gaia} DR2 $G$, $G_\mathrm{BP}$, and $B_\mathrm{RP}$; Johnson $V$ and $B$; and WISE $W1$ and $W2$), and the \texttt{SME} spectroscopic values ($T_\mathrm{eff}$, log$g$, and [Fe/H]) as priors. The posteriors are sampled using the \texttt{MultiNest} \citep{Feroz2009, Feroz2019} sampling algorithm.
    
    All values for the stellar radius and mass are reported in \autoref{tab:mass_radius}. We include values from the TIC and \textit{Gaia} DR2, as well as the typical mass and radius for an F7V dwarf for reference \citep{Cox2000}. The \texttt{SpecMatch-Emp} fit also derives a stellar radius, which we couple to the calibration equations from \citet{Torres2010} to infer a surface gravity of log$g = 4.14 \pm 0.07$ dex and a stellar mass of $1.25 \pm 0.09$ $M_\odot$. All values are in good agreement with each other. We adopt the \texttt{ARIADNE} radius ($R_* = 1.316 \pm 0.019$ $R_\odot$) and the \texttt{isochrones} mass ($M_* = 1.21 \pm 0.02$ $M_\odot$) as the stellar parameters to be used in the global fit, and report all adopted physical, photometric, and kinematic properties of TOI-1670 in \autoref{tab:star_params}.
    
    \begin{deluxetable}{ccc}[!t]
    \setlength{\tabcolsep}{8pt}
    \tablecaption{Stellar mass and radius of TOI-1670 derived from different methods. \label{tab:mass_radius}}
    \tablehead{\colhead{Method} & \colhead{$M_* (M_\odot)$} & \colhead{$R_* (R_\odot)$}}
    \startdata
    \texttt{ARIADNE}\tablenotemark{\footnotesize{a}} & $1.16 \pm 0.16$ & $1.316 \pm 0.019$ \\
    \texttt{isochrones} & $1.21 \pm 0.02$ & $1.316 \pm 0.007$ \\
    \texttt{SpecMatch-Emp}+Torres\tablenotemark{\footnotesize{b}} & $1.25 \pm 0.09$ & $1.57 \pm 0.18$ \\
    \texttt{TIC}\tablenotemark{\footnotesize{c}} & $1.25 \pm 0.18$ & $1.312 \pm 0.057$ \\
    \textit{Gaia} DR2\tablenotemark{\footnotesize{d}} & $\cdots$ & $1.38_{-0.07}^{+0.08}$ \\
    Typical F7V dwarf\tablenotemark{\footnotesize{e}} & 1.21 & 1.32 \\
    \hline
    Adopted & $1.21 \pm 0.02$ & $1.316 \pm 0.019$ \\
    \enddata
    \tablecomments{$^{(a)}$Mass calculated using derived radius and log$g$. $^{(b)}$Mass calculated using \texttt{SpecMatch-Emp} parameters and calibration equations from \citet{Torres2010}. $^{(c)}$\citet{Stassun2019}. $^{(d)}$\citet{Gaia2018}. $^{(e)}$\citet{Cox2000}.}
    \vspace{-7.5mm}
    \end{deluxetable}

    \begin{deluxetable}{ccc}
    \setlength{\tabcolsep}{6pt}
    \tablecaption{Adopted Physical, Photometric, and Kinematic Properties of TOI-1670. \label{tab:star_params}}
    \tablehead{\colhead{Parameter} & \colhead{Value} & \colhead{Source}}
    \startdata
        TIC ID & 441739020 & 1 \\
        TOI ID & 1670 & 1 \\
        \textit{Gaia} ID & 1651911084230149248 & 2 \\
        2MASS ID & J17160415+7209402 & 3 \\
        \textit{Gaia} $\alpha$ (J2000.0) & 17:16:04.16 & 2 \\
        \textit{Gaia} $\delta$ (J2000.0) & +72:09:40.17 & 2 \\
        \textit{Gaia} Epoch & 2015.5 & 2 \\
        \textit{Gaia} Parallax (mas) & $5.92\pm0.02$ & 2 \\
        Distance (pc) & $165.72_{-0.38}^{+0.32}$ & 4 \\
        \textit{Gaia} $\mu_\alpha$cos$\delta$ (mas yr$^{-1}$) & $-6.09\pm0.05$ & 2 \\
        \textit{Gaia} $\mu_\delta$ (mas yr$^{-1}$) & $5.154\pm0.05$ & 2 \\
        \hline
        $B$ (mag) & $10.43\pm0.03$ & 5 \\
        $V$ (mag) & $9.89\pm0.03$ & 5 \\
        $T$ (mag) & $9.423\pm0.0061$ & 1 \\
        $G$ (mag) & $9.8232\pm0.0004$ & 2 \\
        $G_{RP}$ (mag) & $9.4145\pm0.0014$ & 2 \\
        $G_{BP}$ (mag) & $10.0747\pm0.0010$ & 2 \\
        $J$ (mag) & $8.97\pm0.02$ & 3 \\
        $H$ (mag) & $8.75\pm0.03$ & 3 \\
        $K_s$ (mag) & $8.724\pm0.02$ & 3 \\
        $W1$ (mag) & $8.689\pm0.023$ & 6 \\
        $W2$ (mag) & $8.702\pm0.020$ & 6 \\
        \hline
        $T_\mathrm{eff}$ (K) & $6170 \pm 61$ & This work \\
        log $g$ (g cm$^{-3}$) & $4.29 \pm 0.11$ & This work \\
        {[}Fe/H{]} (dex) & $0.09 \pm 0.07$ & This work \\
        $v$sin$i$ (km s$^{-1}$) & $9.2 \pm 0.6$ & This work \\
        $M_*$ ($M_\odot$) & $1.21 \pm 0.02$ & This work \\
        $R_*$ ($R_\odot$) & $1.316 \pm 0.019$ & This work \\
        $\rho_*$ (g cm$^{-3}$) & $0.752 \pm 0.036$ & This work \\
        Age (Gyr) & $2.53 \pm 0.43$ & This work \\
        $A_V$ (mag) & $0.010 \pm 0.006$ & This work
    \enddata
    \tablerefs{(1) \citet{Stassun2019}, (2) \citet{Gaia2018}, (3) \citet{Cutri2003}, (4) \citet{Bailer-Jones2021}, (5) \citet{Hog2000}, (6) \citet{Cutri2014}}
    \vspace{-7.5mm}
    \end{deluxetable}

\subsection{Stellar Activity}
\label{sec:activity}

    Stellar activity in the form of rotationally modulated starspots and granulation can both mimic and mask the signals of planets in light curves \citep{Llama2015, Llama2016} and radial velocities \citep[e.g.,][]{Figueira2013}. Thus, prior to running the global model fit, we first examine if stellar activity significantly influences the light curve and RV time series of TOI-1670. We measured a low average value of log$R^\prime_\mathrm{HK} = -4.93 \pm 0.01$ from the HARPS-N spectra, which suggests that TOI-1670 is a quiet star not dominated by stellar activity \citep{Mamajek2008}. The \tess light curve prior to detrending also does not exhibit any significant rotation or activity-induced variability.
    
\begin{figure}[!t]
	    \centerline{\includegraphics[width=1.0\linewidth]{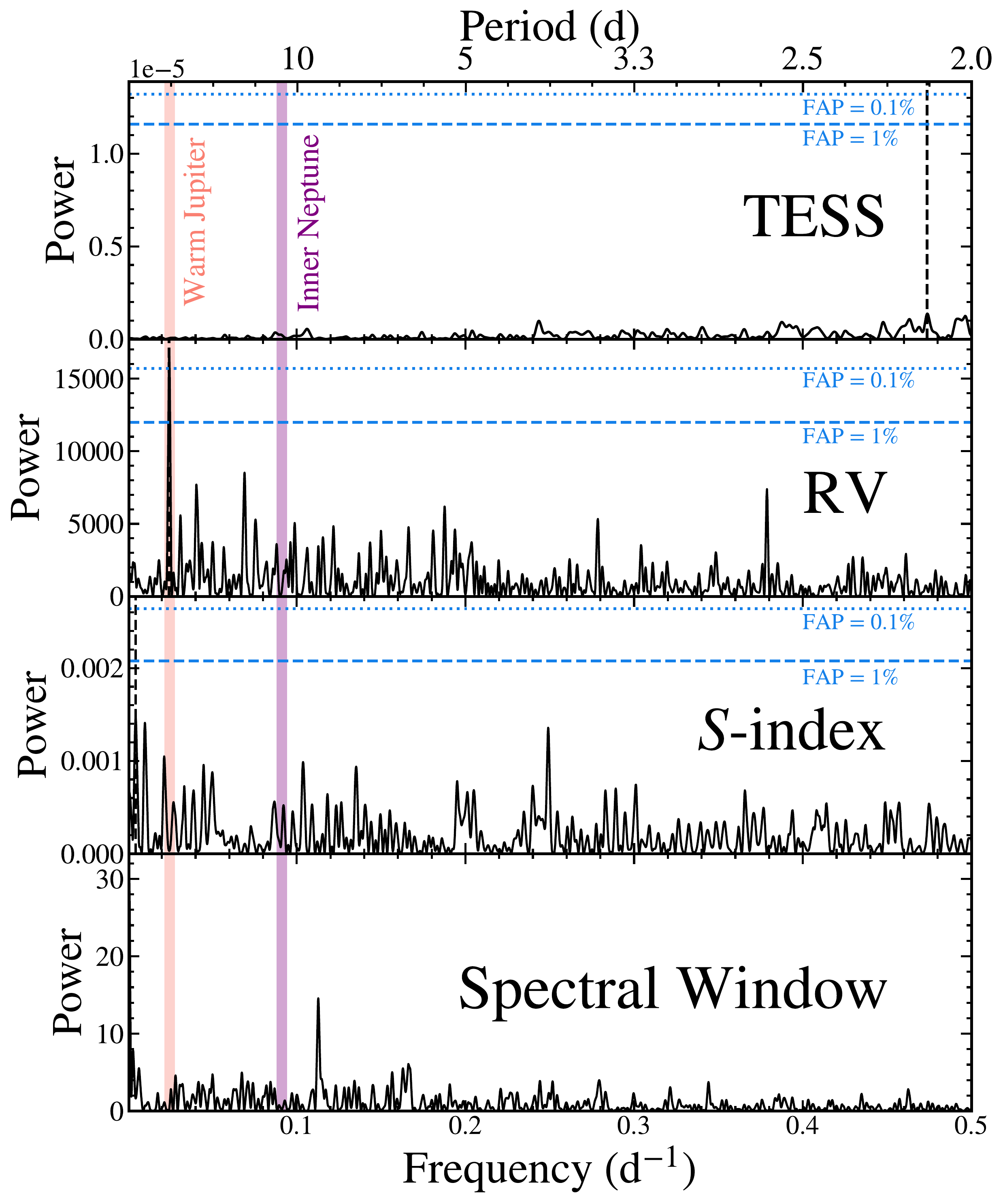}}
	    \caption{Lomb-Scargle periodograms of the undetrended \tess photometry after removing transit signals (top), combined Tull coud\'e and HARPS-N RVs after subtracting the systemic velocities (second), combined Mt. Wilson \textit{S}-index from Tull coud\'e and HARPS-N spectra (third), and spectral window function (bottom). 1\% and 0.1\% FAP (blue dashed and dotted, respectively) lines are calculated using bootstrap resampling. The purple and pink vertical lines are the $\sim$10.9 and $\sim$40.7 d planetary signals determined by the \tess SPOC, respectively. The highest peak of each periodogram is shown as a dashed black vertical line. There are no counterpart peaks in the periodogram of the \textit{S}-index that correspond to the $\sim$40.7 day signal seen in the periodogram of the RV, and no peaks rise above the 1\% FAP threshold. No peaks are visible in the spectral window function periodogram. The only significant period ($<$0.1\% FAP) in the RVs is at 41.1 d, consistent with the WJ (TOI-1670 c).}
	    \label{fig:activity}
    \end{figure}

    A common statistical tool used to detect periodic signals in unevenly sampled time series data is the Lomb-Scargle periodogram \citep{Lomb1976, Scargle1982}. We utilize this algorithm to search for periodicity in both the \tess photometry and RV activity indicators to distinguish stellar activity-based signals from those induced by planetary motion. We compute the Generalized Lomb-Scargle periodograms \citep[GLS;][]{Zechmeister2009} for the ``undetrended'' PDCSAP light curve with the transit events removed, the RVs, the aforementioned activity indicators, and the spectral window function over the frequency range $0.0005-0.5$ d$^{-1}$ ($2-2000$ days) in \autoref{fig:activity}. The GLS power thresholds corresponding to false alarm probability (FAP) levels of 1\% and 0.1\% computed via a bootstrap approach are shown as dotted blue lines \citep{Kuerster1997}. The GLS periodogram for the RVs was computed for the combined HARPS-N and Tull coud\'e data, after subtracting the systematic velocity offsets as reported in \autoref{tab:model_params}. The periodogram of the \tess photometry has very low power with no peaks that have significance higher than the 1\% FAP level. This is consistent with the flat nature of the undetrended PDCSAP light curve and indicates that TOI-1670 does not have a large starspot coverage fraction. The strongest signal in the periodogram of the RVs is at the $\sim$40.7 day orbital period of TOI-1670 c, which has a FAP $<$ 0.1\%. This peak has no counterparts in the periodograms of the activity indices, which would be the case if that signal originated from stellar activity. Activity signals can also appear at the frequency of the stellar rotation period. Using the stellar radius and $v$sin$i$, we can place a lower limit of $P_\mathrm{rot} \geq 7.2$ d ($f \leq 0.138$ d$^{-1}$). No significant peaks are visible in the GLS periodogram of the $S$-indices at this frequency.
    
\subsection{Statistical Validation of TOI-1670 b}
    \label{sec:confim}
    
    Although TOI-1670 b is not significantly detected in the RV dataset, our model is able to place an upper limit on its mass. The $3\sigma$ upper limit of the fitted RV semi-amplitude is 14.5 m s$^{-1}$, which corresponds to an upper limit of 0.18 $M_\mathrm{Jup}$ for TOI-1670 b assuming an eccentricity of 0.

    An estimate of the RV precision required to robustly detect TOI-1670 b can be made from a predicted mass inferred from its radius. Inputting the stellar and planetary parameters measured in \autoref{sec:analysis} into a probabilistic mass-radius relation using the open software package \texttt{forecaster} \citep{Chen2017} yields a mass estimate of $5.2_{-2.0}^{+4.0}$ $M_\earth$ for TOI-1670 b. Assuming a circular orbit, this corresponds to an RV semi-amplitude of $\sim$1.3 m s$^{-1}$. Robustly detecting an RV signal at this level requires an instrument precision at the 1 m s$^{-1}$ level and a well-behaved star.
    
    We can exclude false positive scenarios to support TOI-1670 b as a likely planet using follow-up observations. From \textit{Gaia} EDR3, we note that TOI-1670 has zero excess astrometric noise and a re-normalised unit weight error (RUWE) of 1.07, indicating that the single-star model is a good fit to the astrometric solution \citep{Gaia2018, Lindegren2018}. From our radial velocities, we find that the overall RV variability is $<$110 m s$^{-1}$ from OES, $<$54 m s$^{-1}$ from TRES, $<$34 m s$^{-1}$ from the Tull coud\'e, and $<$16 m s$^{-1}$ from HARPS-N, all of which robustly exclude an eclipsing binary scenario for the host star.
    
    Finally, we use \texttt{TRICERATOPS} \citep{Giacalone2021} to statistically evaluate the probability of possible false positive scenarios involving nearby contaminant stars, including background eclipsing binaries. \texttt{TRICERATOPS} is a Bayesian tool for validating transiting planet candidates by modeling and calculating the probability of different scenarios that produce transit-like light curves. Based on the lack of a close stellar companion from \textit{Gaia} astrometry, our high-resolution imaging, and our RVs, we omit the optional false positive calculations for the eclipsing binary and unresolved stellar companion scenarios in the \texttt{TRICERATOPS} code.\footnote{\citet{Giacalone2021} note that using follow-up observations to rule out unresolved stellar companion scenarios produces similar results for both \texttt{TRICERATOPS} and the target validation code \texttt{vespa} \citep{Morton2015a, Morton2016}.} \texttt{TRICERATOPS} returns a false positive probability (the total probability of a false positive scenario involving the primary star) of $<$0.015 and a nearby false positive probability (the sum of all false positive probabilities for scenarios involving nearby stars) of $<$10$^{-2}$. The RV confirmation of the outer coplanar transiting WJ further supports the planetary nature of TOI-1670 b as multi-planet systems are unlikely to be false positives \citep{Lissauer2012, Rowe2014}.

\subsection{Joint Modeling of RVs \& Photometry}\label{sec:modeling}

    We perform a multi-planet global fit to the available RV and transit observations of TOI-1670 using the \texttt{pyaneti} modeling suite \citep{Barragan2019}. As the planetary Doppler signals are not recovered at a significant level in the TRES and Ond\v rejov spectra, we only use the 49 Tull coud\'e and the 8 HARPS-N radial velocities in the modeling. We limit the light curve data to photometry spanning four full transit duration before and after all transit events of TOI-1670 b and c to improve computation efficiency; this results in a total of 24772 photometric points. These regions are colored purple and pink in \autoref{fig:lightcurve} for TOI 1670 b and c, respectively.

    We simultaneously fit the Keplerian orbit and \tess light curve for 8 parameters: orbital period ($P$), central time of transit ($T_0$), RV semi-amplitude ($K$), transit impact parameter ($b$), planetary-to-stellar radius ($R_\mathrm{p}/R_*$), scaled semi-major axis ($a/R_*$), and parameterized forms of eccentricity and argument of periastron ($\sqrt{e}\:\mathrm{sin}\:\omega$ and $\sqrt{e}\:\mathrm{cos}\:\omega$). This last parametrization by \citet{Anderson2011} is used because the eccentricity posterior distribution for orbits with low $e$ and broad $\omega$ is poorly sampled by Markov chains \citep[e.g.,][]{Lucy1971, Ford2006, Wang2011}. By defining $e$ and $\omega$ in a polar form, we avoid truncating the posterior distribution at zero and impose a uniform prior on $e$. We also adopt the parametrization of $b$ as defined by \citet{Winn2010},
    \begin{equation}
        b = \frac{a\:\mathrm{cos}i_*}{R_*}\left(\frac{1 - e^2}{1+e\mathrm{sin}\omega_*}\right),
    \end{equation}
    where $i_*$ is the stellar inclination, in order to impose priors that exclude non-transiting orbits $(b > 1 + \frac{R_p}{R_*})$.

    We set narrow uniform priors on both orbital period and time of transit based on visual inspection of the light curve and the SPOC preliminary parameters. For the inner sub-Neptune the ranges are $T_{0, b} = \mathcal{U}(1721.92, 1721.99)$ in units of (BJD\textsubscript{TDB} – 2457000) d, $P_b = \mathcal{U}(10.980, 10.988)$ d, and $K_b = \mathcal{U}(0.0, 10.0)$ m s$^{-1}$. For the outer Jupiter the ranges are $T_{0, c} = \mathcal{U}(1750.82, 1750.92)$ in units of (BJD\textsubscript{TDB} – 2457000) d, $P_c = \mathcal{U}(40.7485, 40.7505)$ d, and $K_c = \mathcal{U}(10.0, 100.0)$ m s$^{-1}$. The stellar mass and radius are also free parameters with Gaussian priors of $M_* = \mathcal{N}(1.215, 0.023)\; M_\odot$ and $R_* = \mathcal{N}(1.316, 0.019)\; R_\odot$.\footnote{$\mathcal{U}$ and $\mathcal{N}$ refer to the uniform and normal distributions, respectively, where the latter is defined as $\mathcal{N}(\mu, \sigma)$.} These parameters are further constrained by the stellar mean density, which is affected by $P$ and $a/R_*$ \citep{Seager2003, Winn2010}. We assumed a quadratic limb-darkening law following the equations from \citet{Mandel2002}, who define the linear and quadratic coefficients as $u_1$ and $u_2$, respectively. The parameterization of $q_1 = (u_1 + u_2)^2$ and $q_2 = 0.5 u_1 (u_1 + u_2)^{-1}$ from \citet{Kipping2013} is adopted. We set broad uniform priors for all other parameters and report them in \autoref{tab:model_params}. A ``jitter'' term is added to the radial velocities to account for any systematic and astrophysical variance not reported in the observational uncertainties.\footnote{We also fit for a noise term in the photometry and find a value two orders of magnitude less than the typical uncertainty with no appreciable change in other model parameters. We choose not to include this term in the final joint model fit.}

    Posterior distributions of fitted and derived parameters were sampled using an MCMC Metropolis-Hasting algorithm following the description by \citet{Sharma2017} as implemented by \texttt{pyaneti}. The distributions were sampled using 50 chains for 10000 iterations with a thinning factor of 10. The convergence of each chain was determined with the Gelman-Rubin diagnostic test \citep{Gelman1992}.
    
\begin{figure}[!t]
        \centering
        \includegraphics[width=1.0\linewidth]{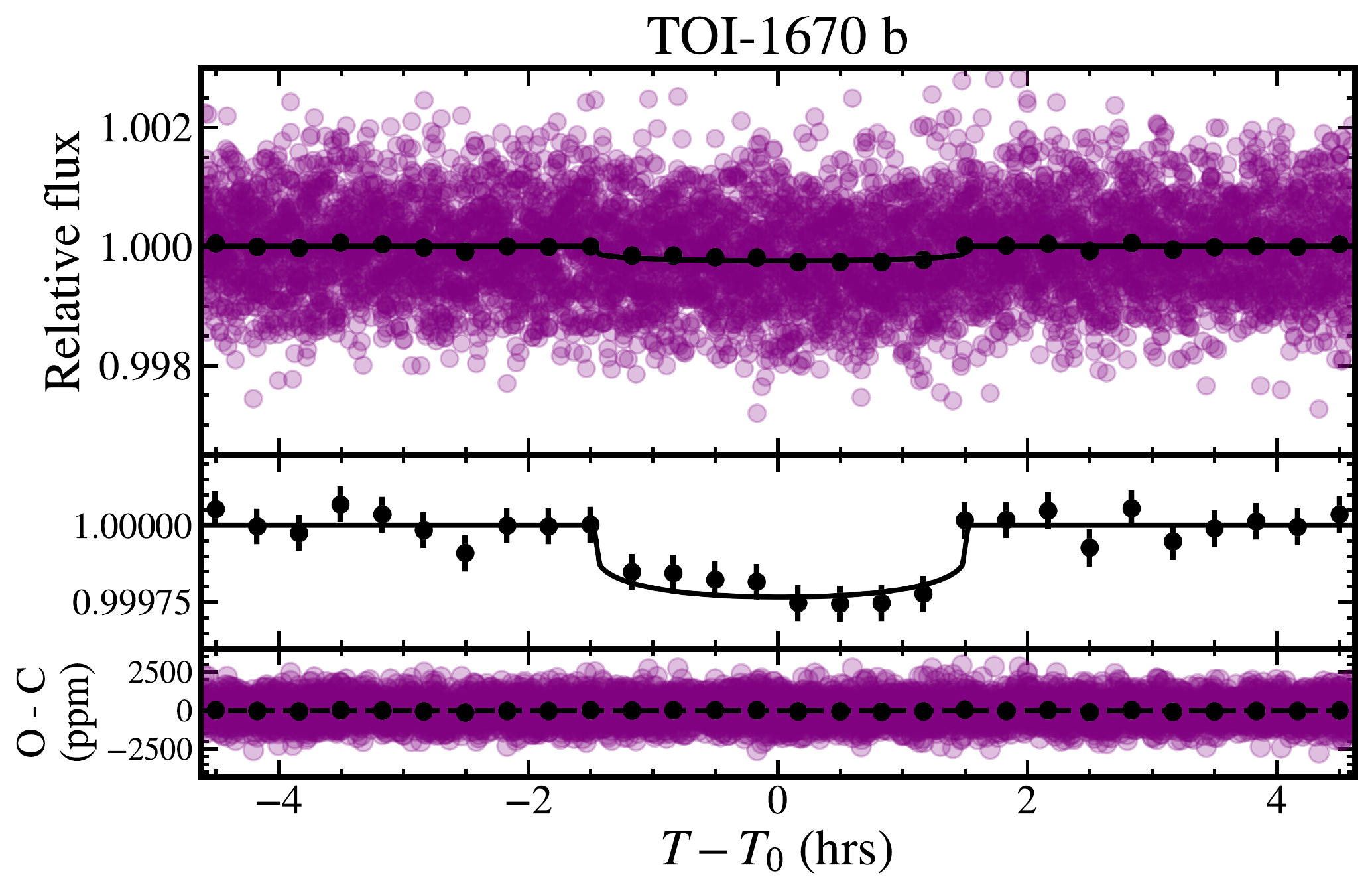}
	    \caption{Transit light curve folded to the orbital period of TOI-1670 b. The \tess photometry are shown in purple and the solid black line is the best-fitting transit model. The black circles are the photometric data binned over 20-minute intervals. The middle panel zooms in on the best-fit transit model and binned data points. The fit residuals are shown in the lower panel.}
	    \label{fig:lc_period_nep}
    \end{figure}

\begin{figure}[!t]
        \centering
        \includegraphics[width=1.0\linewidth]{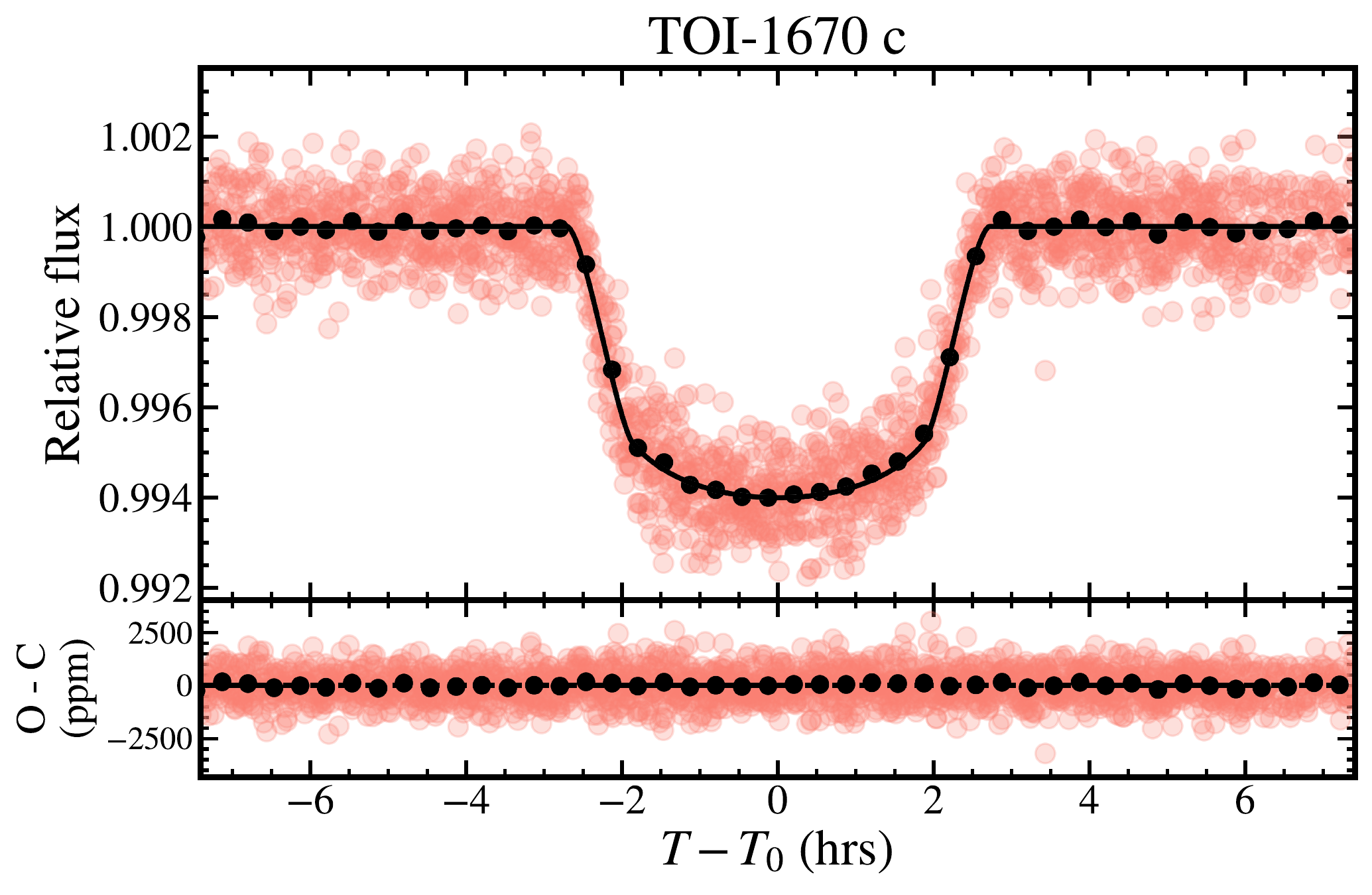}
	    \caption{Transit light curve folded to the orbital period of TOI-1670 c. The \tess photometry are shown in pink and the solid black line is the best-fitting transit model. Black points are the photometric data binned over 20-minute intervals.}
	    \label{fig:lc_period_jup}
    \end{figure}

\begin{figure}[!tp]
        \centering
        \includegraphics[width=1.0\linewidth]{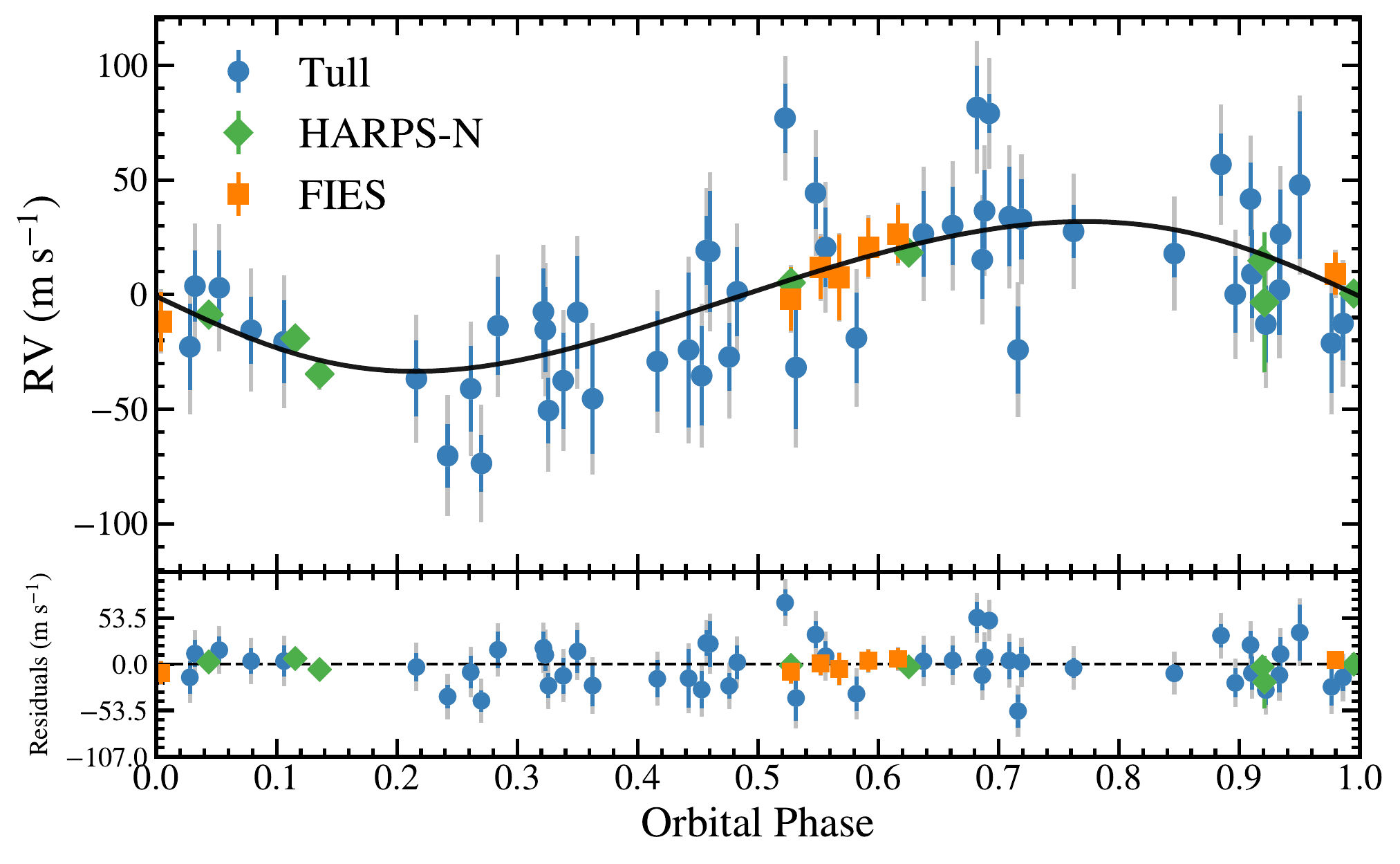}
	    \caption{RV curve of TOI-1670 phase folded to the orbital period of the WJ, TOI-1670 c, with the contributions of the inner companion removed. The different color points denote the different spectrographs, and the best-fitting RV model is shown in the solid black line. The fit residuals are shown in the lower panel. The colored error bars are nominal RV errors and the grey error bars include the systematic jitter term.}
	    \label{fig:rvs}
    \end{figure}

    Using the \tess photometry and RVs, we jointly model the transits of TOI-1670 b and c and the RV curve of TOI-1670 c using the priors as previously described.\footnote{We also consider less complex models in \autoref{sec:appendix_model_selection}. We ultimately choose to apply a full global fit to robustly assess parameter uncertainties.} The posterior values of the fitted and derived system parameters from \texttt{pyaneti} for TOI-1670 are given in \autoref{tab:model_params}. The best-fitting phased \tess light curves for TOI-1670 b and c and RV model for TOI-1670 c are plotted in Figures \ref{fig:lc_period_nep}--\ref{fig:rvs}. \autoref{fig:cornerplot} in \autoref{sec:appendix_corners} displays the posterior distributions for fitted parameters. We find a mass, radius, and density of TOI-1670 c of $M_\mathrm{c} = 0.63_{-0.08}^{+0.09}$ $M_\mathrm{Jup}$, $R_\mathrm{c} = 0.987_{-0.025}^{+0.025}$ $R_\mathrm{Jup}$, and $\rho_\mathrm{c} = 0.81_{-0.11}^{+0.13}$ g cm$^{-3}$, respectively. For TOI-1670 b, we find a radius of $2.06_{-0.15}^{+0.19}$ $R_\earth$ and a $3\sigma$ mass upper limit of $M_\mathrm{b} < 0.13$ $M_\mathrm{Jup}$.
    
    \begin{deluxetable*}{lllcc}[!ht]
    \renewcommand{\arraystretch}{1.2}
    \tablecaption{Priors and Posteriors on Global System Parameters of TOI-1670 b and c.
    \label{tab:model_params}}
    \tablehead{\lhead{Parameter} & \multicolumn{2}{c}{Adopted Prior} & \multicolumn{2}{c}{Posterior Values}}
    \startdata
        \textbf{Fitted Parameters} & \multicolumn{1}{c}{$b$} & \multicolumn{1}{c}{$c$} & \multicolumn{1}{c}{$b$} & \multicolumn{1}{c}{$c$} \\
        $T_0$ (BJD\textsubscript{TDB}$-2457000$) & $\mathcal{U}(1721.92, 1721.99)$ & $\mathcal{U}(1750.82, 1750.92)$ & $1721.9423_{-0.0062}^{+0.0071}$ & $1750.88286_{-0.00083}^{+0.00085}$ \\
        $P$ (days) & $\mathcal{U}(10.980, 10.988)$ & $\mathcal{U}(40.7485, 40.7505)$ & $10.98462_{-0.00051}^{+0.00046}$ & $40.74976_{-0.00021}^{+0.000022}$ \\
        $K$ (m s$^{-1}$) & $\mathcal{U} (0.0, 10.0)$ & $\mathcal{U} (10.0, 100)$ & $4.6_{-3.0}^{+3.3}$ & $32.7_{-4.3}^{+4.7}$ \\
        $b$ & $\mathcal{U}(0.0, 1.0)$ & $\mathcal{U}(0.0, 1.0)$ & $0.61_{-0.37}^{+0.22}$ & $0.76_{-0.04}^{+0.02}$ \\
        $a/R_*$ & $\mathcal{U}(1.1, 20.0)$ & $\mathcal{U}(1.1, 50.0)$ & $16.88_{-0.27}^{+0.27}$ & $40.68_{-0.66}^{+0.66}$ \\
        $R_\mathrm{p}/R_*$ & $\mathcal{U}(0.0, 0.05)$ & $\mathcal{U}(0.0, 0.15)$ & $0.014_{-0.001}^{+0.001}$ & $0.077_{-0.002}^{+0.002}$ \\
        $\sqrt{e}$ sin $\omega$ & $\mathcal{U}(-1.0, 1.0)$ & $\mathcal{U}(-1.0, 1.0)$ & $0.18_{-0.44}^{+0.36}$ & $0.27_{-0.10}^{+0.08}$ \\
        $\sqrt{e}$ cos $\omega$ & $\mathcal{U}(-1.0, 1.0)$ & $\mathcal{U}(-1.0, 1.0)$ & $-0.63_{-0.21}^{+0.62}$ & $-0.07_{-0.13}^{+0.14}$ \\
        \hline
        \textbf{Derived Parameters} & & & & \\
        $M_p$ & \multicolumn{1}{c}{$\cdots$} & \multicolumn{1}{c}{$\cdots$} & $13.8_{-8.7}^{+9.5}$ $M_\earth$ & $0.63_{-0.08}^{+0.09}$ $M_\mathrm{Jup}$ \\
        $R_p$ & \multicolumn{1}{c}{$\cdots$} & \multicolumn{1}{c}{$\cdots$} & $2.06_{-0.15}^{+0.19}$ $R_\earth$ & $0.987_{-0.025}^{+0.025}$ $R_\mathrm{Jup}$ \\
        $\rho_p$ (g cm$^{-3}$) & \multicolumn{1}{c}{$\cdots$} & \multicolumn{1}{c}{$\cdots$} & $8.6_{-5.6}^{+6.9}$ & $0.81_{-0.11}^{+0.13}$ \\
        $e$ & \multicolumn{1}{c}{$\cdots$} & \multicolumn{1}{c}{$\cdots$} & $0.59_{-0.26}^{+0.17}$ & $0.09_{-0.04}^{+0.05}$ \\
        $\omega$ (deg) & \multicolumn{1}{c}{$\cdots$} & \multicolumn{1}{c}{$\cdots$} & $163.6_{-53.7}^{+41.7}$ & $105.5_{-29.4}^{+28.6}$ \\
        $i$ (deg) & \multicolumn{1}{c}{$\cdots$} & \multicolumn{1}{c}{$\cdots$} & $86.87_{-1.07}^{+1.16}$ & $88.84_{-0.04}^{+0.04}$ \\
        $a$ (AU) & \multicolumn{1}{c}{$\cdots$} & \multicolumn{1}{c}{$\cdots$} & $0.103_{-0.002}^{+0.002}$ & $0.249_{-0.005}^{+0.005}$ \\
        $T_{14}$ (hrs) & \multicolumn{1}{c}{$\cdots$} & \multicolumn{1}{c}{$\cdots$} & $2.80_{-0.19}^{+0.16}$ & $5.40_{-0.05}^{+0.06}$ \\
        $T_{eq}$ (K) & \multicolumn{1}{c}{$\cdots$} & \multicolumn{1}{c}{$\cdots$} & $1062_{-13}^{+14}$ & $684_{-9}^{+9}$ \\
        \hline
        \textbf{Additional Parameters} & & & & \\
        $M_*$ (from scaled parameters) ($M_\odot$) & \multicolumn{2}{l}{\hspace{2cm}$\mathcal{N}(1.215, 0.023)$} & $1.218_{-0.076}^{+0.081}$ & $1.239_{-0.079}^{+0.084}$ \\
        $\rho_*$ (from transit) (g cm$^{-3}$) & \multicolumn{2}{l}{\hspace{2cm}$\cdots$} & $0.754_{-0.035}^{+0.037}$ & $0.767_{-0.037}^{+0.038}$ \\
        $q_1$ & \multicolumn{2}{l}{\hspace{2cm}$\mathcal{U} (0.0, 1.0)$} & \multicolumn{2}{c}{$0.35_{-0.11}^{+0.19}$} \\
        $q_2$ & \multicolumn{2}{l}{\hspace{2cm}$\mathcal{U} (0.0, 1.0)$} & \multicolumn{2}{c}{$0.32_{-0.23}^{+0.39}$} \\
        $\gamma_\mathrm{Tull}$ (km s$^{-1}$) & \multicolumn{2}{l}{\hspace{2cm}$\mathcal{U}(-3.6849, -3.3299)$} & \multicolumn{2}{c}{$-3.5132_{-0.0043}^{+0.0042}$} \\
        $\gamma_\mathrm{FIES}$ (km s$^{-1}$) & \multicolumn{2}{l}{\hspace{2cm}$\mathcal{U}(-0.1217, 0.1139)$} & \multicolumn{2}{c}{$-0.0108_{-0.0061}^{+0.0061}$} \\
        $\gamma_\mathrm{HARPS-N}$ (km s$^{-1}$) & \multicolumn{2}{l}{\hspace{2cm}$\mathcal{U}(-11.6134, -11.3602)$} & \multicolumn{2}{c}{$-11.4805_{-0.0027}^{+0.0025}$} \\
        RV jitter (Tull) (m s$^{-1}$) & \multicolumn{2}{l}{\hspace{2cm}$\cdots$} & \multicolumn{2}{c}{$22.5_{-3.6}^{+3.9}$} \\
        RV jitter (FIES) (m s$^{-1}$) & \multicolumn{2}{l}{\hspace{2cm}$\cdots$} & \multicolumn{2}{c}{$5.3_{-3.8}^{+7.2}$} \\
        RV jitter (HARPS-N) (m s$^{-1}$) & \multicolumn{2}{l}{\hspace{2cm}$\cdots$} & \multicolumn{2}{c}{$4.3_{-2.8}^{+4.5}$} \\
    \enddata
    \end{deluxetable*}
    
    \section{\textbf{Discussion}} \label{sec:discuss}
    
    \begin{figure*}[!t]
        \centerline{\includegraphics[width=1.0\linewidth]{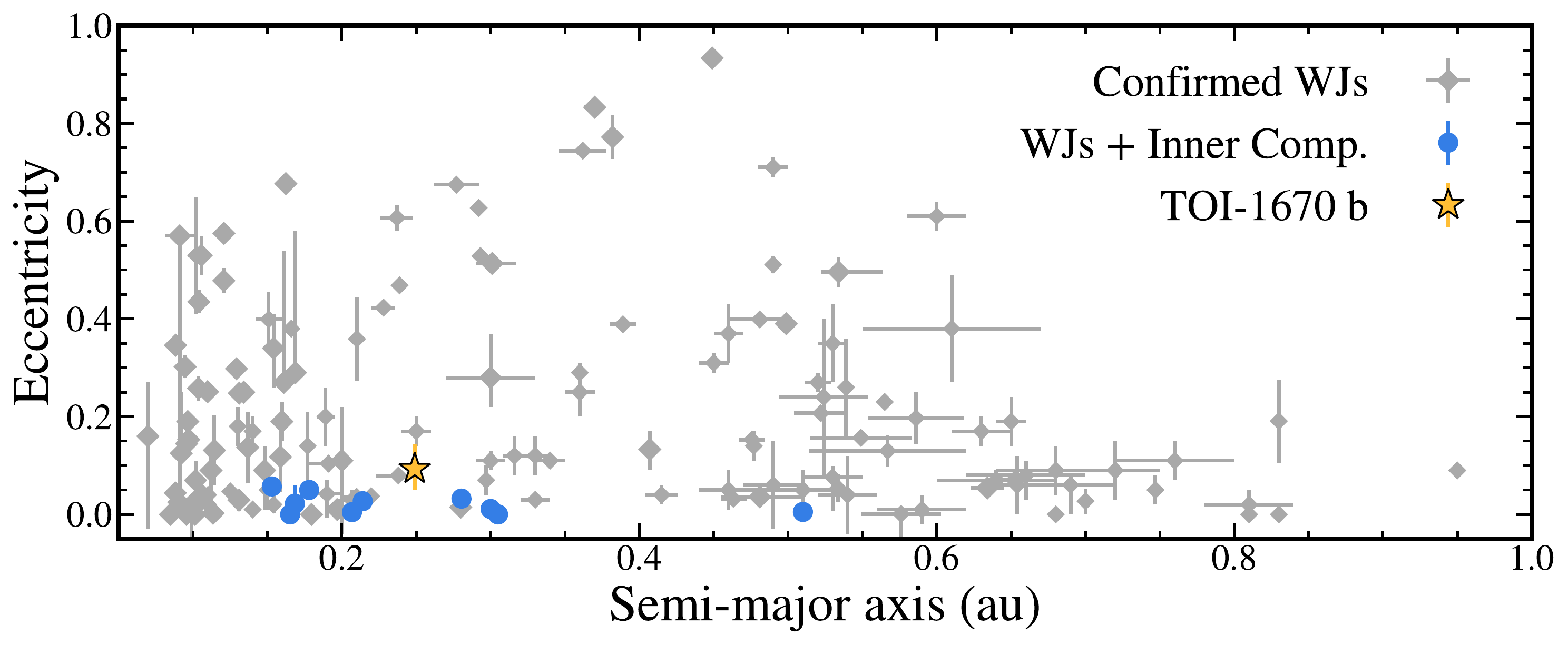}}
	    \caption{Eccentricity as a function of semi-major axis for confirmed WJ systems. Multi-planet systems with an outer WJ and a smaller, inner companion are shown in blue circles whereas all other WJs are plotted in grey diamonds. TOI-1670 is denoted with the yellow star. Systems with a similar configuration as TOI-1670 all have low eccentricities, suggesting a gentle formation pathway. Planet orbital parameters are assembled from exoplanet.eu \citep{Schneider2011} and the NASA Exoplanet Archive \citep{Akeson2013, nasa_exo_archive} as of June 2021.}
	    \label{fig:other_wjs}
    \end{figure*}
    
    \begin{deluxetable*}{cccrccrcccc}[!t]
    \setlength{\tabcolsep}{8pt}
    \tablecaption{Confirmed systems with an outer warm giant planet and at least one smaller inner companion. \label{tab:wj+friends}}
    \tablehead{\colhead{System} & \colhead{$M_\mathrm{in}$} & \colhead{$R_\mathrm{in}$} & \colhead{$P_\mathrm{in}$} &
    \colhead{$M_\mathrm{out}$} &
    \colhead{$R_\mathrm{out}$} & \colhead{$P_\mathrm{out}$} & \colhead{$e_\mathrm{out}$} &
    \colhead{$\sigma_e$\tablenotemark{\footnotesize{b}}} & \colhead{$P_\mathrm{out}$/$P_\mathrm{in}$} & \colhead{Ref.} \\ \colhead{} & \colhead{$(M_\mathrm{Jup})$} & \colhead{$(R_\mathrm{Jup})$} & \colhead{(d)} & \colhead{$(M_\mathrm{Jup})$} & \colhead{$(R_\mathrm{Jup})$} & \colhead{(d)} & \colhead{} & \colhead{} & \colhead{} & \colhead{}}
    \startdata
    Kepler-89 & $<$0.05 & 0.385 & 10.42 & 0.33 & 1.005 & 22.34 & 0.022 & $0.038$ & 2.14 & 1, 2 \\
    TOI-216 & 0.06 & 0.714 & 17.16 & 0.56 & 0.901 & 34.53 & 0.0046 & $^{+0.0027}_{-0.0012}$ & 2.01 & 3, 4 \\
    Kepler-117 & 0.09 & 0.719 & 18.80 & 1.84 & 1.101 & 50.79 & 0.0323 & $0.0033$ & 2.70 & 5, 6 \\
    Kepler-30 & 0.03 & 0.348 & 29.22 & 1.69 & 1.097 & 60.32 & 0.011 & $0.001$ & 2.06 & 7, 8 \\
    HIP 57274 & 0.04\tablenotemark{\footnotesize{b}} & $\cdots$ & 8.14 & 0.41\tablenotemark{\footnotesize{b}} & $\cdots$ & 32.03 & 0.05 & $0.03$ & 3.94 & 9 \\
    GJ 876 & 0.76\tablenotemark{\footnotesize{c}} & $\cdots$ & 30.13 & 2.39\tablenotemark{\footnotesize{c}} & $\cdots$ & 61.08 & 0.027 & $0.002$ & 2.03 & 10 \\
    K2-290 & $<$0.07 & 0.273 & 9.21 & 0.77 & 1.006 & 48.37 & 0 (fixed) & $<$0.241 & 5.25 & 11 \\
    Kepler-56 & 0.07 & 0.581 & 10.50 & 0.57 & 0.874 & 21.40 & 0.00 & $0.01$ & 2.04 & 12, 13 \\
    Kepler-88 & 0.03 & 0.307 & 10.92 & 0.67 & $\cdots$ & 22.26 & 0.0572 & $0.0005$ & 2.04 & 14 \\
    Kepler-289 & 0.01 & 0.239 & 66.06 & 0.42 & 1.034 & 125.85 & 0.005 & $0.015$ & 1.91 & 15 \\
    TOI-1670 & $<$0.13 & 0.184 & 10.98 & 0.63 & 0.987 & 40.75 & 0.09 & $_{-0.04}^{+0.05}$ & 3.71 & This work
    \enddata
    \tablecomments{$^{(a)}$1$\sigma$ uncertainties or 3$\sigma$ upper limit on $e_\mathrm{out}$. $^{(b)}$Minimum mass, $M_p$sin$i$. $^{(c)}$Mass determined assuming coplanar model with fixed inclinations.}
    \tablerefs{(1) \citet{Hirano2012}, (2) \citet{Weiss2013}, (3) \citet{Dawson2019}, (4) \citet{Dawson2021}, (5) \citet{Rowe2014}, (6) \citet{Bruno2015}, (7) \citet{Sanchis-Ojeda2012}, (8) \citet{Panichi2018}, (9) \citet{Fischer2012}, (10) \citet{Trifonov2018}, (11) \citet{Hjorth2019}, (12) \citet{Huber2013}, (13) \citet{Otor2016}, (14) \citet{Weiss2020}, (15) \citet{Schmitt2014}.}
    \end{deluxetable*}
    
    The existence of WJ systems hosting one or more smaller coplanar inner companions such as TOI-1670 is inconsistent with dynamical migration routes. During high-eccentricity tidal migration, multiple close-in planets would likely interact with each other and potentially lead to ejections or collisions \citep[e.g.,][]{Rasio1996, Chatterjee2008, Mustill2015}. Similarly, planet-planet scattering and Von Zeipel–Lidov–Kozai interactions require an outer companion \citep{Veras2005, Anderson2017}. Multi-planet systems hosting a WJ with low eccentricity represent another type of system that experienced comparatively gentle dynamical histories such as inward disk migration or \textit{in situ} formation.
    
    TOI-1670 joins a handful of confirmed systems with an outer warm, giant exoplanet ($M_p > 0.25$ $M_\mathrm{Jup}$, $10<P<200$ d) and at least one inner, smaller companion (see \autoref{tab:wj+friends}). \autoref{fig:other_wjs} shows the eccentricity versus semi-major axis of confirmed WJs. WJs in all 11 systems (including TOI-1670) with similar configurations have low eccentricities, whereas the eccentricities of other WJs without known inner companions are widely distributed. This divergence further suggests that this group of TOI-1670-like systems may have formed and migrated along a similar evolutionary pathway.
    
    One way to disentangle whether disk migration or \textit{in situ} formation plays the dominant role in sculpting these multi-planet systems is by examining their period ratios in search of near mean-motion resonances (MMRs). Disk migration is expected to efficiently capture giant planets into MMRs close to small integer period ratios such as 2:1, 3:1, 3:2, and 4:3 \citep[e.g.,][]{Goldreich1980, Lee2001, Armitage2010, Winn2015}. \textit{In situ} formation can also create planets in orbital resonances, either coincidentally or by eccentricity damping via interactions with the protoplanetary or planetesimal disk \citep{Dawson2016, Morrison2020}. In this formation scenario there should be a population of systems that congregate at or near these different integer ratios.
    
    Within the sample of 11 known systems that have a giant planet with a small inner companion, 8 WJs (73\%) are in or near a 2:1 or 3:1 resonance with the inner planet (\autoref{tab:wj+friends}). With a period ratio of 3.7, TOI-1670 joins two other systems, K2-290 and HIP 57274, that have non-MMR orbital period ratios greater than 3. The planets in these systems may have formed \textit{in situ} or migrated inward together. Alternatively, that the planets in these systems are not locked in an MMR could also indicate that they formed independently and did not migrate together or became unstable over time \citep{Pichierri2020, Petit2020, Izidoro2021}. This may hint at a division within this small class of WJs in multi-planet systems in which some migrate into place via disk migration (those with integer period ratios) while others formed where we see them today or experienced further dynamical interaction later in their lifetime. This hypothesis can be further investigated by increasing the number of warm giant planets with smaller inner companions and examining population trends within this sample.
    
    \acknowledgments
    We thank Benjamin Tofflemire, Daniel Krolikowski, Michael Gully-Santiago, and Erik Petigura for insightful discussions on the generalized Lomb-Scargle periodogram, gas giant occurrence rate, and light curve analysis.
    
    Q.H.T. and B.P.B. acknowledge the support from a NASA FINESST grant (80NSSC20K1554). This work benefited from involvement in ExoExplorers, which is sponsored by the Exoplanets Program Analysis Group (ExoPAG) and NASA’s Exoplanet Exploration Program Office (ExEP). B.P.B. acknowledges support from the National Science Foundation grant AST-1909209 and NASA Exoplanet Research Program grant 20-XRP20$\_$2-0119.
    
    C.M.P. and M.F. gratefully acknowledge the support of the Swedish National Space Agency (DNR 65/19 and 177/19). J.K. gratefully acknowledge the support of the Swedish National Space Agency (SNSA; DNR 2020-00104). P.K., M.S., J.S., and R.K. acknowledge the financial support of the Inter-transfer grant no LTT-20015. M.K. acknowledges support from ESAs PEA4000127913. M.E. acknowledges the support of the DFG priority program SPP 1992 ``Exploring the Diversity of Extrasolar Planets'' (HA 3279/12-1). Funding for the Stellar Astrophysics Centre is provided by The Danish National Research Foundation (Grant agreement no.: DNRF106). D.G. and L.M.S. gratefully acknowledge financial support from the Cassa di Risparmio di Torino (CRT) foundation under Grant No. 2018.2323 ``Gaseous or rocky? Unveiling the nature of small worlds''. This work is partly supported by JSPS KAKENHI Grant Number JP20K14518 and SATELLITE Research from Astrobiology Center (AB022006).

    Funding for the TESS mission is provided by NASA's Science Mission Directorate. This research has made use of the Exoplanet Follow-up Observation Program website, which is operated by the California Institute of Technology, under contract with the National Aeronautics and Space Administration under the Exoplanet Exploration Program. We acknowledge the use of public TESS data from pipelines at the TESS Science Office and at the TESS Science Processing Operations Center. Resources supporting this work were provided by the NASA High-End Computing (HEC) Program through the NASA Advanced Supercomputing (NAS) Division at Ames Research Center for the production of the SPOC data products. This paper includes data collected by the TESS mission that are publicly available from the Mikulski Archive for Space Telescopes (MAST).
    
    Observations in the paper made use of the NN-EXPLORE Exoplanet and Stellar Speckle Imager (NESSI). NESSI was funded by the NASA Exoplanet Exploration Program and the NASA Ames Research Center. NESSI was built at the Ames Research Center by Steve B. Howell, Nic Scott, Elliott P. Horch, and Emmett Quigley. The authors are honored to be permitted to conduct observations on Iolkam Du'ag (Kitt Peak), a mountain within the Tohono O'odham Nation with particular significance to the Tohono O'odham people.
    
    Observations in the paper made use of the High-Resolution Imaging instrument ‘Alopeke. ‘Alopeke was funded by the NASA Exoplanet Exploration Program and built at the NASA Ames Research Center by Steve B. Howell, Nic Scott, Elliott P. Horch, and Emmett Quigley. ‘Alopeke was mounted on the Gemini North telescope of the international Gemini Observatory, a program of NSF’s NOIRLab, which is managed by the Association of Universities for Research in Astronomy (AURA) under a cooperative agreement with the National Science Foundation on behalf of the Gemini partnership: the National Science Foundation (United States), National Research Council (Canada), Agencia Nacional de Investigación y Desarrollo (Chile), Ministerio de Ciencia, Tecnología e Innovación (Argentina), Ministério da Ciência, Tecnologia, Inovações e Comunicações (Brazil), and Korea Astronomy and Space Science Institute (Republic of Korea).
    
    This work was enabled by observations made from the Gemini North telescope, located within the Maunakea Science Reserve and adjacent to the summit of Maunakea. We are grateful for the privilege of observing the Universe from a place that is unique in both its astronomical quality and its cultural significance.
    
    Based on observations made with the Nordic Optical Telescope, operated by the Nordic Optical Telescope Scientific Association at the Observatorio del Roque de los Muchachos, La Palma, Spain, of the Instituto de Astrofisica de Canarias under program 59-210.
    
    \facilities{McDonald Observatory: 2.7-m Harlan J. Smith Telescope (Tull coud\'e), Roque de los Muchachos Observatory: 3.58-m Telescopio Nazionale Galileo (HARPS-N), 2.56-m Nordic Optical Telescope (FIES), \textit{TESS}, FLWO: 1.5-m (TRES), Ond\v rejov Observatory: 2-m Perek Telescope (OES), Gemini Observatory: 8.1-m Gemini North telescope (`Alopeke).}
    
    \software{\texttt{pyaneti} \citep{Barragan2019}, \texttt{lightkurve} \citep{Lightkurve2018}, \texttt{Austral} \citep{Endl2000}, \texttt{astropy} \citep{Astropy2018}, \texttt{matplotlib} \citep{Hunter4160265}, \texttt{SME} \citep{Valenti1996, Valenti2005, Piskunov2017}, \texttt{SpecMatch-Emp} \citep{Yee2017}, \texttt{ARIADNE} \citep{Vines2021}, \texttt{forecaster} \citep{Chen2017}, \texttt{isochrones} \citep{Morton2015b}, \texttt{TRICERATOPS} \citep{Giacalone2021}.}

    \restartappendixnumbering

    \appendix
    \section{RV measurements} \label{sec:appendix_rvs}
    \autoref{tab:recon_RVs} records the reconnaissance RV measurements for TOI-1670. \autoref{tab:RVs} lists the relative Tull coud\'e and HARPS-N precise RVs used in the joint global orbit fit of TOI-1670. See \autoref{sec:observations} for details.
    
    \begin{deluxetable*}{cccc}[!htp]
    \setlength{\tabcolsep}{12pt}
    \tablecaption{Reconnaissance RV Measurements \label{tab:recon_RVs}}
    \tablehead{\colhead{BJD\textsubscript{TDB}} & \colhead{RV (m s$^{-1}$)} & \colhead{$\sigma_\mathrm{RV}$ (m s$^{-1}$)} & \colhead{Instrument}}
    \startdata
    2458882.039318 & 0.0 & 53.0 & TRES \\
    2458900.020153 & 47.3  &  51.0 & TRES \\
    2458915.012011 & 24.0  &  55.2 & TRES \\
    2458917.976858 & 154.6 &  50.7 & TRES \\
    2458925.016694 & 120.9 &  59.5 & TRES \\
    2459037.890277 &  47.8 &  53.0 & TRES \\
    2458891.503545280 & $-44487.8$ & 65.5 & OES \\
    2458894.510498418 & $-44361.4$ & 235.5 & OES \\
    2458930.563753183 & $-44463.7$ & 148.9 & OES \\
    2458931.473755364 & $-44129.3$ & 133.1 & OES \\
    2458931.513916946 & $-44314.2$ & 86.1 & OES \\
    2458936.492023797 & $-44303.4$ & 78.8 & OES \\
    2458937.487545487 & $-44271.6$ & 96.2 & OES \\
    2458945.510337937 & $-44305.1$ & 79.2 & OES \\
    2458947.486183500 & $-44194.3$ & 142.3 & OES \\
    2458953.506692706 & $-44245.2$ & 144.2 & OES \\
    2458956.498833757 & $-44480.6$ & 101.3 & OES \\
    2458956.517247843 & $-44259.6$ & 89.7 & OES \\
    2458957.559932472 & $-44282.0$ & 127.1 & OES \\
    2458959.481333630 & $-44108.1$ & 164.9 & OES \\
    2458959.523497538 & $-44053.2$ & 90.4 & OES \\
    2458959.503590371 & $-44141.2$ & 68.8 & OES \\
    2458960.545742497 & $-44363.6$ & 77.3 & OES \\
    2458961.436750593 & $-44160.5$ & 143.6 & OES \\
    2458963.472548647 & $-44261.4$ & 85.7 & OES \\
    2458962.515129209 & $-44262.5$ & 59.6 & OES \\
    2458964.553762859 & $-44301.1$ & 84.5 & OES \\
    2458967.598704441 & $-44164.8$ & 61.7 & OES \\
    2458976.525095871 & $-44269.5$ & 102.3 & OES \\
    2458989.502480241 & $-44023.8$ & 108.6 & OES \\
    2458991.474925697 & $-44320.5$ & 124.4 & OES \\
    2459002.550308453 & $-44299.1$ & 114.2 & OES \\
    2459067.550946345 & $-44228.8$ & 162.3 & OES \\
    2459071.496445452 & $-44133.1$ & 125.1 & OES \\
    2459074.552383421 & $-44246.4$ & 119.8 & OES \\
    2459100.533604675 & $-44300.5$ & 145.7 & OES \\
    2459101.536958157 & $-44393.2$ & 96.2 & OES \\
    2459104.526104830 & $-44470.6$ & 65.0 & OES \\
    \enddata
    \end{deluxetable*}
    
    \clearpage
    
    \startlongtable
    \begin{deluxetable*}{cccccccc}
    \setlength{\tabcolsep}{10pt}
    \tablecaption{Relative RV Measurements Used in Orbit Fit \label{tab:RVs}}
    \tablehead{\colhead{BJD\textsubscript{TDB}} & \colhead{RV} & \colhead{$\sigma_\mathrm{RV}$} & \colhead{Instrument} & \colhead{S-index} & \colhead{$\sigma_\mathrm{S-index}$} & \colhead{BIS} & \colhead{FWHM} \\
    \colhead{(d)} & \colhead{(m s$^{-1}$)} & \colhead{(m s $^{-1}$)} & \colhead{} & \colhead{} & \colhead{} & \colhead{(m s$^{-1}$)} & \colhead{(km s$^{-1}$)}}
    \startdata
    2459073.581564 & $-11469.3$ & 3.7 & HARPS-N & $0.158$ & $0.001$ & 24.3 & 13.764 \\
    2459082.412035 & $-11513.4$ & 5.6 & HARPS-N & $0.161$ & $0.002$ & 31.2 & 13.816 \\
    2459098.371706 & $-11475.3$ & 3.1 & HARPS-N & $0.160$ & $0.001$ & 16.8 & 13.780 \\
    2459102.360849 & $-11460.2$ & 3.6 & HARPS-N & $0.159$ & $0.001$ & 16.0 & 13.733 \\
    2459114.404597 & $-11481.9$ & 30.5 & HARPS-N & $\cdots$ & $\cdots$ & 74.8 & 13.744 \\
    2459117.430022 & $-11482.9$ & 4.1 & HARPS-N & $0.164$ & $0.002$ & 27.0 & 13.806 \\
    2459119.408423 & $-11490.8$ & 3.5 & HARPS-N & $0.158$ & $0.001$ & 26.2 & 13.743 \\
    2459122.336070 & $-11498.2$ & 2.9 & HARPS-N & $0.160$ & $0.001$ & 6.9 & 13.758 \\
    \hline
    2458994.555398 & $0.0$ & $9.2$ & FIES & $\cdots$ & $\cdots$ & 10.2 & 18.393 \\
    2458995.551629 & $-21.7$ & $13.0$ & FIES & $\cdots$ & $\cdots$ & 41.2 & 18.509 \\
    2459018.504159 & $-5.7$ & $18.6$ & FIES & $\cdots$ & $\cdots$ & 32.4 & 18.365 \\
    2459019.496162 & $3.7$ & $12.8$ & FIES & $\cdots$ & $\cdots$ & 22.7 & 18.411 \\
    2459020.497833 & $13.9$ & $12.7$ & FIES & $\cdots$ & $\cdots$ & 7.4 & 18.458 \\
    2459098.364441 & $-12.7$ & $13.8$ & FIES & $\cdots$ & $\cdots$ & 16.8 & 18.394 \\
    2459099.369294 & $1.9$ & $13.6$ & FIES & $\cdots$ & $\cdots$ & 24.0 & 18.442 \\
    \hline
    2458950.979568 & $-3502.7$ & $19.3$ & Tull coud\'e & $0.20$ & $0.03$ & $\cdots$ & $\cdots$ \\
    2458951.909478 & $-3510.9$ & $19.5$ & Tull coud\'e & $0.21$ & $0.03$ & $\cdots$ & $\cdots$ \\
    2458957.843670 & $-3527.2$ & $14.6$ & Tull coud\'e & $0.21$ & $0.03$ & $\cdots$ & $\cdots$ \\
    2458958.958375 & $-3531.9$ & $18.2$ & Tull coud\'e & $0.19$ & $0.03$ & $\cdots$ & $\cdots$ \\
    2458982.835596 & $-3432.2$ & $08.5$ & Tull coud\'e & $0.20$ & $0.03$ & $\cdots$ & $\cdots$ \\
    2458983.927587 & $-3478.8$ & $17.5$ & Tull coud\'e & $0.19$ & $0.03$ & $\cdots$ & $\cdots$ \\
    2458994.813377 & $-3524.2$ & $16.1$ & Tull coud\'e & $0.19$ & $0.02$ & $\cdots$ & $\cdots$ \\
    2459047.698935 & $-3524.8$ & $21.3$ & Tull coud\'e & $0.20$ & $0.03$ & $\cdots$ & $\cdots$ \\
    2459054.754109 & $-3493.7$ & $15.1$ & Tull coud\'e & $0.19$ & $0.03$ & $\cdots$ & $\cdots$ \\
    2459054.873732 & $-3494.0$ & $26.4$ & Tull coud\'e & $0.19$ & $0.03$ & $\cdots$ & $\cdots$ \\
    2459055.802778 & $-3510.8$ & $19.5$ & Tull coud\'e & $0.19$ & $0.02$ & $\cdots$ & $\cdots$ \\
    2459072.664836 & $-3512.6$ & $16.9$ & Tull coud\'e & $0.19$ & $0.02$ & $\cdots$ & $\cdots$ \\
    2459073.701460 & $-3530.7$ & $16.8$ & Tull coud\'e & $0.19$ & $0.02$ & $\cdots$ & $\cdots$ \\
    2459090.659723 & $-3548.9$ & $21.0$ & Tull coud\'e & $0.19$ & $0.02$ & $\cdots$ & $\cdots$ \\
    2459091.655351 & $-3556.7$ & $24.2$ & Tull coud\'e & $0.20$ & $0.02$ & $\cdots$ & $\cdots$ \\
    2459104.674112 & $-3429.9$ & $18.2$ & Tull coud\'e & $\cdots$ & $\cdots$ & $\cdots$ & $\cdots$ \\
    2459115.597718 & $-3463.8$ & $32.1$ & Tull coud\'e & $0.20$ & $0.03$ & $\cdots$ & $\cdots$ \\
    2459116.673732 & $-3534.0$ & $21.7$ & Tull coud\'e & $0.19$ & $0.02$ & $\cdots$ & $\cdots$ \\
    2459134.594159 & $-3540.6$ & $21.8$ & Tull coud\'e & $0.18$ & $0.02$ & $\cdots$ & $\cdots$ \\
    2459135.654423 & $-3535.4$ & $33.8$ & Tull coud\'e & $0.20$ & $0.02$ & $\cdots$ & $\cdots$ \\
    2459143.603909 & $-3485.6$ & $18.6$ & Tull coud\'e & $0.20$ & $0.02$ & $\cdots$ & $\cdots$ \\
    2459144.595144 & $-3481.6$ & $17.1$ & Tull coud\'e & $0.18$ & $0.03$ & $\cdots$ & $\cdots$ \\
    2459145.600434 & $-3496.1$ & $17.2$ & Tull coud\'e & $0.19$ & $0.02$ & $\cdots$ & $\cdots$ \\
    2459171.552605 & $-3528.1$ & $18.6$ & Tull coud\'e & $0.20$ & $0.02$ & $\cdots$ & $\cdots$ \\
    2459228.014988 & $-3486.3$ & $21.8$ & Tull coud\'e & $0.20$ & $0.03$ & $\cdots$ & $\cdots$ \\
    2459241.016167 & $-3536.2$ & $18.8$ & Tull coud\'e & $0.20$ & $0.03$ & $\cdots$ & $\cdots$ \\
    2459242.012350 & $-3509.5$ & $16.2$ & Tull coud\'e & $0.19$ & $0.02$ & $\cdots$ & $\cdots$ \\
    2459242.012350 & $-3509.5$ & $16.2$ & Tull coud\'e & $0.18$ & $0.02$ & $\cdots$ & $\cdots$ \\
    2459270.939084 & $-3486.9$ & $11.5$ & Tull coud\'e & $0.19$ & $0.02$ & $\cdots$ & $\cdots$ \\
    2459275.930375 & $-3455.1$ & $13.6$ & Tull coud\'e & $\cdots$ & $\cdots$ & $\cdots$ & $\cdots$ \\
    2459276.929526 & $-3469.8$ & $16.0$ & Tull coud\'e & $0.18$ & $0.02$ & $\cdots$ & $\cdots$ \\
    2459277.949543 & $-3484.9$ & $19.6$ & Tull coud\'e & $0.18$ & $0.03$ & $\cdots$ & $\cdots$ \\
    2459281.939450 & $-3510.8$ & $15.5$ & Tull coud\'e & $0.18$ & $0.02$ & $\cdots$ & $\cdots$ \\
    2459293.909641 & $-3570.9$ & $14.4$ & Tull coud\'e & $0.17$ & $0.02$ & $\cdots$ & $\cdots$ \\
    2459294.893907 & $-3523.4$ & $24.7$ & Tull coud\'e & $0.18$ & $0.02$ & $\cdots$ & $\cdots$ \\
    2459301.921161 & $-3434.4$ & $15.1$ & Tull coud\'e & $0.14$ & $0.02$ & $\cdots$ & $\cdots$ \\
    2459302.952236 & $-3467.8$ & $15.7$ & Tull coud\'e & $0.18$ & $0.02$ & $\cdots$ & $\cdots$ \\
    2459309.807855 & $-3535.6$ & $19.0$ & Tull coud\'e & $0.18$ & $0.03$ & $\cdots$ & $\cdots$ \\
    2459339.846070 & $-3548.9$ & $21.8$ & Tull coud\'e & $0.18$ & $0.03$ & $\cdots$ & $\cdots$ \\
    2459340.775733 & $-3539.7$ & $14.8$ & Tull coud\'e & $0.19$ & $0.02$ & $\cdots$ & $\cdots$ \\
    2459355.848218 & $-3493.3$ & $10.4$ & Tull coud\'e & $0.20$ & $0.03$ & $\cdots$ & $\cdots$ \\
    2459372.775403 & $-3554.6$ & $18.7$ & Tull coud\'e & $0.19$ & $0.03$ & $\cdots$ & $\cdots$ \\
    2459383.791707 & $-3545.3$ & $26.6$ & Tull coud\'e & $0.18$ & $0.02$ & $\cdots$ & $\cdots$ \\
    2459384.793846 & $-3491.9$ & $17.4$ & Tull coud\'e & $0.18$ & $0.02$ & $\cdots$ & $\cdots$ \\
    2459385.835820 & $-3530.8$ & $19.8$ & Tull coud\'e & $\cdots$ & $\cdots$ & $\cdots$ & $\cdots$ \\
    2459411.687127 & $-3548.1$ & $16.6$ & Tull coud\'e & $\cdots$ & $\cdots$ & $\cdots$ & $\cdots$ \\
    2459412.744002 & $-3582.4$ & $13.9$ & Tull coud\'e & $\cdots$ & $\cdots$ & $\cdots$ & $\cdots$ \\
    2459454.638202 & $-3584.9$ & $12.5$ & Tull coud\'e & $\cdots$ & $\cdots$ & $\cdots$ & $\cdots$ \\
    2459456.737608 & $-3519.6$ & $18.8$ & Tull coud\'e & $\cdots$ & $\cdots$ & $\cdots$ & $\cdots$ \\
    2459471.673405 & $-3476.9$ & $17.6$ & Tull coud\'e & $\cdots$ & $\cdots$ & $\cdots$ & $\cdots$ \\
    \enddata
    \end{deluxetable*}
    
    \section{Model Complexity and Selection} \label{sec:appendix_model_selection}
    
    Model selection balances the quality of the model fit and the model complexity, or number of parameters. Including extraneous parameters can lead to over-fitting of the data while excluding physically-motivated aspects of the model can result in under-fitting of the data and introduce bias. Our model fit is susceptible to the former scenario when we consider parameters that can not be robustly estimated from the data, such as the mass of the sub-Neptune or the eccentricities of either planet.
    
    Here we assess whether two alternative less complex models are more justified by the data using different model selection criteria. We first compare model fits using only the RV data (excluding the light curve) for a one- and two-planet model, whereby we exclude the inner sub-Neptune in the one-planet fit. In the second comparison, we examine two joint model fits, one where the WJ eccentricity is fixed to zero and another in which the WJ eccentricity is a free parameter. In both cases, the inner sub-Neptune is not modeled with the RVs and its eccentricity is fixed to zero.
    
    Several metrics can be used to establish whether a model is justified by the data. For each fit, \texttt{pyaneti} reports the Bayesian Information Criterion \citep[BIC;][]{Schwarz1978, Raftery1986}, defined as
    \begin{equation}
        \text{BIC} \equiv -2\:\text{ln}\:\mathcal{L} + k\:\text{ln}\:N,
    \end{equation}
    and the Akaike Information Criterion \citep[AIC;][]{Akaike1998}, defined as
    \begin{equation}
        \text{AIC} \equiv -2\:\text{ln}\:\mathcal{L} + 2k.
    \end{equation}
    Here, $\mathcal{L}$ is the model likelihood, $k$ is the number of model parameters, and $N$ is the number of data points used in the fit. We further calculate the AIC corrected for small sample sizes \citep[AIC\textsubscript{c};][]{Sugiura1978, Burnham2004}:
    \begin{equation}
        \text{AIC\textsubscript{c}} = \text{AIC} + \frac{2k(k+1)}{N - k - 1}.
    \end{equation}
    This metric is preferred over the AIC as it can be understood as a relative model likelihood using Akaike weights \citep{Akaike1981, Burnham2004, Liddle2007}, where the weight $w$ for each model $i$ is
    \begin{equation}
        w_i = \frac{e^{-\Delta\text{AIC\textsubscript{c,$i$}}/2}}{\sum_{r=1}^{R}e^{-\Delta\text{AIC\textsubscript{c,$r$}}/2}}.
    \end{equation}
    These metrics take into account the model likelihood while penalizing each additional free parameter. The BIC can be interpreted as a model evidence ratio, such that according to the Jeffreys' scale a BIC difference between two models of $\geq$5 is strong evidence and $\geq$10 is decisive evidence against the model with a higher BIC \citep{Jeffreys1935, Kass1995, Liddle2007}.
    
    We find that both the BIC and Akaike weights criteria favor the one-planet model for the RV data and the WJ circular orbit model in the global joint fits. For the RV-only fits, we find $\mathrm{BIC} = -269.3$, $\mathrm{AIC} = -297.3$, $\mathrm{AIC}_\mathrm{c} = -290.1$, and an Akaike weight of $>$0.99 for the one-planet only model and $\mathrm{BIC} = -254.0$, $\mathrm{AIC} = -288.5$, $\mathrm{AIC}_\mathrm{c} = -277.0$, and an Akaike weight of $<$0.01 for the two-planet model. The BIC comparison ($\Delta\text{BIC} > 15$) and the Akaike weight (99\% relative likelihood) strongly favor the one-planet model fit. In the global joint fits, we find $\mathrm{BIC} = -283106.1$, $\mathrm{AIC} = -283260.4$, $\mathrm{AIC}_\mathrm{c} = -283260.3$, and an Akaike weight of 0.85 for the zero-eccentricity global model and $\mathrm{BIC} = -283086.4$, $\mathrm{AIC} = -283256.9$, $\mathrm{AIC}_\mathrm{c} = -283256.9$, and an Akaike weight of 0.15 for the free eccentricity global model. Both the BIC comparison ($\Delta\text{BIC} > 15$) and the Akaike weights (85\% relative likelihood) suggest that there is strong evidence in favor of the circular orbit model.
    
    Despite these model selection preferences, we find that the final parameter uncertainties are slightly underestimated when compared to the full model fit, which suggests that the less complex models are under-fitting the data. Orbital parameter uncertainties are larger by approximately 10--50\% in the full global model (which includes the smaller, inner planet and the eccentricities of both planets) compared to the more simple, fixed zero-eccentricity joint model. For example, the uncertainties of the orbital period, planet mass, and planet radius of the WJ increased by 10\%, 30\%, and 40\%, respectively. By neglecting the smaller, inner planet and the eccentricity of either planet in order to avoid over-fitting the data, the simpler model artificially underestimates the uncertainties of other model parameters. Ultimately, to more accurately determine parameter uncertainties, we adopt the global simultaneous joint model fit and report upper limits on parameter posteriors when robust detections are not possible.
    
    \section{Posterior distributions of fitted parameters} \label{sec:appendix_corners}
\begin{figure*}[!]
        \centerline{\includegraphics[width=1.0\linewidth]{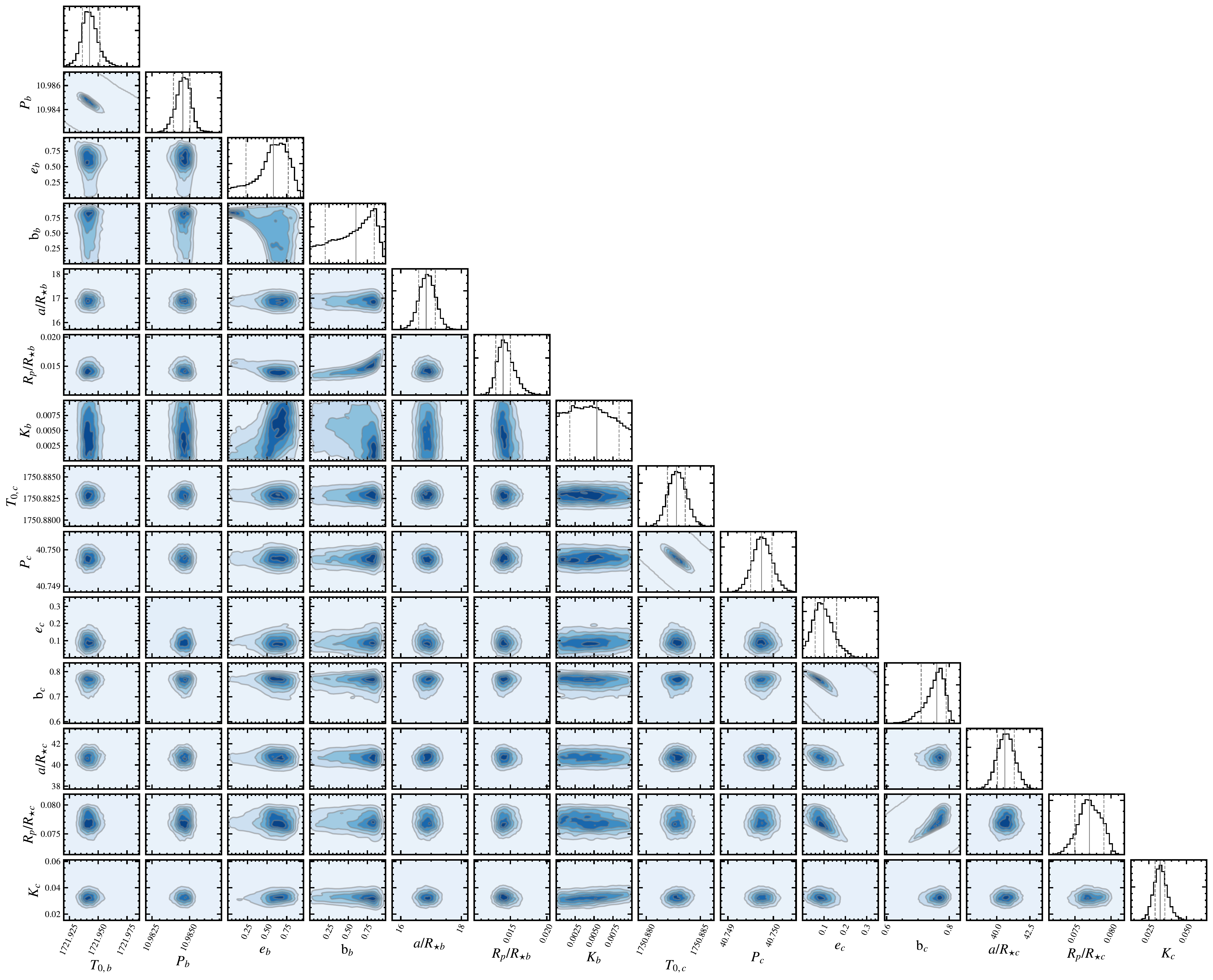}}
	    \caption{Posterior distributions of fitted parameters for TOI-1670 b and c derived from joint global fit of RVs and light curve.}
	    \label{fig:cornerplot}
    \end{figure*}

    \autoref{fig:cornerplot} displays the posterior distributions of fitted parameters from the global joint fit. See \autoref{sec:modeling} for more details.
    
    \clearpage
    
    \bibliographystyle{aasjournal}
    \bibliography{toi-1670}{}

    \end{document}